%% file: article_v1.tex
\RequirePackage{fix-cm}
\documentclass[twocolumn]{svjour3}          % twocolumn
\smartqed  % flush right qed marks, e.g. at end of proof
\usepackage{graphicx}
\usepackage{wrapfig}
\usepackage[table]{xcolor}
\usepackage[left]{eurosym}
\begin{document}
\title{EChO%\thanks{Grants or other notes
}
\subtitle{Exoplanet Characterisation Observatory}
\author{ \textnormal{\mbox{Tinetti G.$^{1}$ (\emph{UCL}),   Beaulieu J.P.$^{2}$ (\emph{IAP}),  Henning T.$^{3}$ (\emph{MPIA}),  
Meyer M.$^{4}$ (\emph{ETH}), Micela G.$^{5}$ (\emph{INAF}), }
 Ribas I.$^{6}$ (\emph{IEEC-CSIC}),   Stam D.$^{7}$ (\emph{SRON}),  Swain M.$^{8}$ (\emph{JPL})    \and \newline 
\mbox{Krause  O. (\emph{MPIA}),  Ollivier M. (\emph{IAS}),  Pace E. (\emph{Un. Firenze}),  Swinyard B. (\emph{UCL-RAL})  \and Aylward   A. (\emph{UCL}),}
   \mbox{van Boekel R.  (\emph{MPIA}),   Coradini A. (\emph{INAF}), Encrenaz T.  (\emph{LESIA, Obs. Paris}), 
        Snellen I.  (\emph{Un. Leiden}),  } 
        \mbox{Zapatero-Osorio M. R. (\emph{CAB}) \and 
Bouwman J.  (\emph{MPIA}),  Cho J. Y-K. (\emph{QMUL}),   Coud\'e du Foresto V. (\emph{LESIA}), 
 } \mbox{Guillot T. (\emph{Obs. Nice}), Lopez-Morales M.  (\emph{IEEC}),  Mueller-Wodarg I. (\emph{Imperial College}),  Palle E. (\emph{IAC}),  } Selsis F. (\emph{Un. Bordeaux}), 
Sozzetti A.  (\emph{INAF})   \and \newline 
%%%%%%%%%%%%%
\mbox{Ade P.A.R. (\emph{Cardiff}),
Achilleos N. (\emph{UCL}),
Adriani A. (\emph{INAF}),
Agnor C. B. (\emph{QMUL}),
Afonso C. (\emph{MPIA}), }
%%%%%%%%
\mbox{Allende Prieto C. (\emph{IAC}),
Bakos G. (\emph{Princeton}),
Barber R. J. (\emph{UCL}),
Barlow M.  (\emph{UCL}),
Bernath P. (\emph{Un. York}), }
%%%%%%%%
\mbox{B\'ezard B. (\emph{LESIA}),
Bord\'e P. (\emph{IAS}),
Brown L.R. (\emph{JPL}),
Cassan A. (\emph{IAP}), 
Cavarroc C. (\emph{IAS}), }
%%%%%%% 
\mbox{Ciaravella A. \emph{INAF}),  
Cockell, C.  \emph{O.U.}),
Coustenis A. (\emph{LESIA}),
Danielski C.  (\emph{UCL}),
Decin L.  (\emph{IvS}), }
%%%%%%%%%
\mbox{De Kok R. (\emph{SRON}),
Demangeon O. (\emph{IAS}),
Deroo P. (\emph{JPL}),
Doel P. (\emph{UCL}),
Drossart P. (\emph{LESIA}), }
%%%%%%%
\mbox{Fletcher L.N. (\emph{Oxford}),
Focardi M. (\emph{Un. Firenze}),
Forget F. (\emph{LMD}),
Fossey S.  (\emph{UCL}),
Fouqu\'e P. (\emph{Obs-MIP}), }
%%%%%%%%%%%
\mbox{Frith J. (\emph{UH}),
Galand M. (\emph{Imperial College}),
Gaulme P. (\emph{IAS}),
Gonz\'alez Hern\'andez J.I. (\emph{IAC}),
 }
%%%%%%%%
\mbox{Grasset O.  (\emph{Un. Nantes}), Grassi D.  (\emph{INAF}),
Grenfell J. L. (\emph{TUB}),
Griffin M. J. (\emph{Cardiff}),
}
%%%%%%%
\mbox{Griffith C. A. (\emph{UoA}),
Gr\"ozinger U. (\emph{MPIA}), 
Guedel M.  (\emph{Un. Vienna}),
Guio P. (\emph{UCL}),
Hainaut O. (\emph{ESO}),
 }
%%%%%%%%%
\mbox{Hargreaves R. (\emph{Un. York}),
Hauschildt P. H. (\emph{HS}),
Heng K. (\emph{ETH}),
Heyrovsky D. (\emph{CU Prague}),
 }
%%%%%
\mbox{Hueso R. (\emph{EHU Bilbao}), Irwin P. (\emph{Oxford}),
Kaltenegger L. (\emph{MPIA}),
Kervella P. (\emph{Paris Obs.}),
Kipping D. (\emph{CfA}),
}
%%%%%%
\mbox{Koskinen T.T. (\emph{UoA}),
Kov\'acs G. (\emph{Konkoy Obs.}),
La Barbera A. (\emph{INAF/IASFP}), 
Lammer H. (\emph{Un. Graz}), }
%%%%%%
\mbox{Lellouch E. (\emph{LESIA}),
Leto G. (\emph{INAF/OACt}),
Lopez Morales M. (\emph{IEEC}),
Lopez Valverde M.A. (\emph{IAA/CSIC}), }
%%%%%%%%%
\mbox{Lopez-Puertas M (\emph{IAA-CSIC}),
Lovis C. (\emph{Obs. Geneve})
Maggio A. (\emph{INAF/OAPa}),
Maillard J.P. (\emph{IAP}),
}
%%%%%%
\mbox{Maldonado Prado J. (\emph{UAM}),
Marquette J.B. (\emph{IAP}),
Martin-Torres F.J. (\emph{CAB}),
Maxted P. (\emph{Un. Keele}),
}
\mbox{Miller S. (\emph{UCL}), Molinari S. (\emph{Un. Firenze}), 
Montes D.  (\emph{UCM}),
Moro-Martin A. (\emph{CAB}),
Moses J.I. (\emph{SSI}),
 }
%%%%%%%%
\mbox{Mousis O.  (\emph{Obs. Besancon}),
Nguyen Tuong N.  (\emph{LESIA}),
Nelson R. (\emph{QMUL}),
Orton G.S. (\emph{JPL}),
Pantin E. (\emph{CEA}),}
%%%%%%%%
\mbox{Pascale E. (\emph{Cardiff}),
Pezzuto S. (\emph{IFSI-INAF}),
Pinfield D. (\emph{UH}),
Poretti E. (\emph{INAF/OAMi}),
Prinja R. (\emph{UCL}), }
%%%%%%
\mbox{Prisinzano L. (\emph{INAF-OAPa}),
Rees J.M. (\emph{LESIA}),
Reiners A. (\emph{IAG}), 
Samuel B. (\emph{IAS}),}
%%%%%%%
\mbox{S\'anchez-Lavega A. (EHU Bilbao),
Sanz Forcada J. (\emph{CAB}),
Sasselov D. (\emph{CfA}),
Savini G. (\emph{UCL}),
Sicardy B. (\emph{LESIA}),}
%%%%%%
\mbox{Smith A. (\emph{MSSL}),
Stixrude L. (\emph{UCL}), 
Strazzulla G. (\emph{INAF/OACt}),
Tennyson J. (\emph{UCL}),
Tessenyi M. (\emph{UCL}), }
%%%%%%%%
\mbox{Vasisht G. (\emph{JPL}),
Vinatier S. (\emph{LESIA}),
Viti S. (\emph{UCL}),
Waldmann~I.~(\emph{UCL}),
White G.J. (\emph{OU-RAL}), }
%%%%%
\mbox{Widemann T. (\emph{LESIA}),
Wordsworth R. (\emph{LMD}),
Yelle R. (\emph{UoA}),
Yung Y. (\emph{Caltech}),
Yurchenko S.N. ~(\emph{UCL}) }
}}
\authorrunning{EChO team} % if too long for running head
\institute{{\bf }\\
$^{1}$Giovanna Tinetti \at
              University College London, London, UK \\
              \email{g.tinetti@ucl.ac.uk}           %  \\
           \and
           $^{2}$Jean Philippe Beaulieu \at
             Institut d'Astrophysique de Paris, Paris, France \\
             \email{beaulieu@iap.fr} 
                       \and
           $^{3}$Thomas Henning \at
             Max Planck Institut fur  Astronomie, Heidelberg, Germany \\
             \email{henning@mpia.de} 
                       \and
           $^{4}$Michael Meyer \at
             Eidgenossische Technische Hochschule Zurich, Switzerland \\
             \email{mmeyer@phys.ethz.ch} 
                       \and
           $^{5}$Giusi Micela \at
             INAF, Osservatorio di Palermo, Italy \\
             \email{giusi@astropa.inaf.it} 
                       \and
                       $^{6}$Ignasi Ribas \at
             Institut de Ciencies de l'Espai, Barcelona, Spain \\
             \email{iribas@ieec.uab.es} 
                       \and
           $^{7}$Daphne Stam \at
             SRON Netherlands Institute for Space Research, Utrecht   \\
             \email{d.m.stam@sron.nl}   
              \and
           $^{8}$Mark Swain \at
             Jet Propulsion Laboratory, Pasadena, US \\
             \email{swain@s383.jpl.nasa.gov}   
}
\date{}
\maketitle
\newpage

\begin{abstract}
A dedicated mission to investigate exoplanetary atmospheres represents a major milestone 
in our quest to understand our place in the universe by placing our Solar 
System in context and by addressing the suitability of planets for the 
presence of life.  EChO -- the Exoplanet Characterisation Observatory 
-- is a mission concept specifically geared for  this purpose.

EChO will provide simultaneous, multi-wavelength
spectroscopic observations on a stable platform that will allow very
long exposures. The use of passive cooling, few moving parts and well
established technology gives a low-risk and potentially long-lived
mission.
EChO will build on observations by Hubble, Spitzer and
ground-based telescopes, which discovered the first molecules and
atoms in exoplanetary atmospheres. However, EChO's configuration and
specifications are designed to study a number of systems in a
consistent manner that will eliminate the ambiguities  
affecting prior observations.  EChO will simultaneously observe a broad
enough spectral region --
from the visible to the mid-infrared -- to constrain from one single spectrum the temperature
structure of the atmosphere, the abundances of the major carbon and
oxygen bearing species, the expected photochemically-produced species
and magnetospheric signatures. The spectral range and resolution 
are tailored to separate bands belonging to up to 30
molecules and retrieve the composition and temperature structure
of planetary atmospheres. 

The target list for EChO includes planets ranging from
Jupiter-sized with equilibrium temperatures $T_{eq}$ up to
2000 K, to those of a few Earth masses, with $T_{eq}$
$\sim$300~K.  The list will include planets with no Solar System
analog, such as the recently discovered planets GJ1214b,
whose density lies between that of terrestrial
and gaseous planets, or the rocky-iron planet
55 Cnc~e, with day-side temperature close to 3000~K.
As the number of detected
exoplanets is growing rapidly each year, and the mass and radius of those
detected steadily decreases, the target
list will be constantly adjusted to include the most interesting
systems.  

We have baselined a
dispersive spectrograph design covering continuously the 0.4--16\,$\mu$m
spectral range in 6 channels (1 in the visible, 5 in the InfraRed), which allows the spectral
resolution to be adapted  from several tens
to several hundreds, depending on the target brightness.  The instrument will be mounted behind a 1.5 m class
telescope, passively cooled to 50~K, with the instrument structure and
optics passively cooled to $\sim$45~K.  
EChO will be placed in a grand halo orbit around L2.  This orbit, in combination with an optimised thermal
shield design, provides a highly stable thermal environment and a high degree of visibility of
the sky   to  observe repeatedly several
tens of targets over the year.
Both the baseline and alternative designs have been
evaluated and no critical items with Technology Readiness Level (TRL) less than 4 to 5 have been
identified.  We have also undertaken a first-order cost and
development plan analysis and find that EChO is easily compatible with
the ESA M-class mission framework.
\end{abstract}
\begin{figure}[h]
%\begin{minipage}{0.54\linewidth}
\centering
\includegraphics[width=\linewidth]{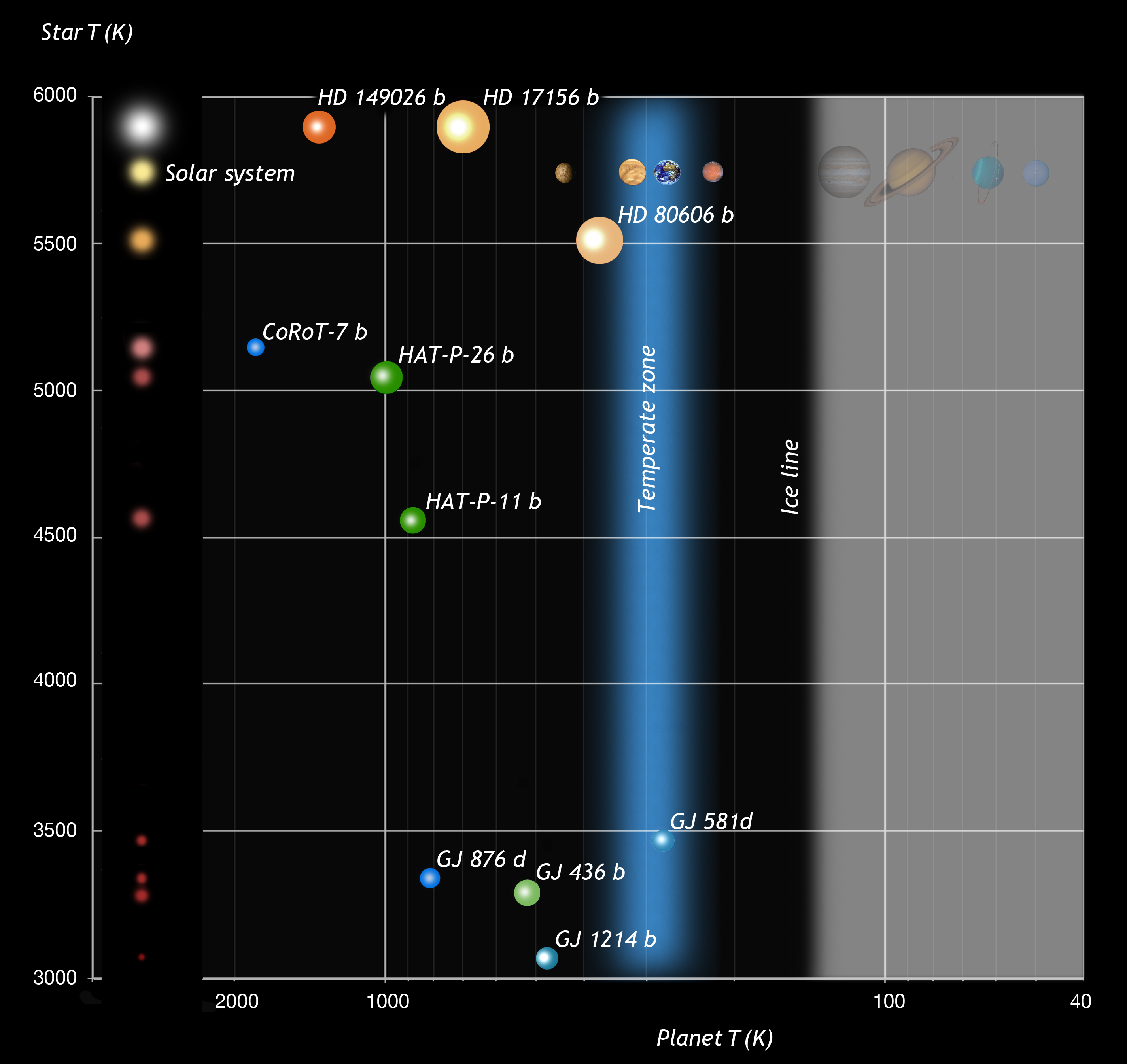}
%\end{minipage}\hfill 
\caption{EChO will expand the scope of planetary science beyond our Solar System, by providing a portfolio of exoplanet spectra under a wide gamut of physical and chemical conditions. The observed chemical composition  largely depends on the planet's thermal structure, which in turn depends on the planet's orbital distance and metallicity, and the host star's luminosity and stellar type. The planetary mass determines the planet's ability to retain an atmosphere. The range of planets and stellar environments explored by EChO extends to the temperate zone and includes gas-giants, Neptunes and super-Earths. It is already populated by  $\sim$200 known transiting objects, and the number of sources is expected to increase exponentially until the launch date, thanks to the current exoplanet discovery programs.}
\label{mastar}
\end{figure} 
\input{scientific_objectives}

\input{observational_strategy}

%\input{scientific_exploitation}
\input{synergy_othermissions}
\input{targets}

%\input{mission_profile}
%%%%%%%%%qui
%%%\input{mission_concept_marc}
%%%%%%%%%%
\input{payload}

%\input{system_requirements}
%%%%\input{science_operations_archiving}
%\input{technology_development_requirements}
\input{conclusions}

 \section*{Ackowledgements}
 We thank Florence Henry, Silvain Cnudde, CNES/PASO, Astrium GmbH Germany, AIM Infrarot-Module GmbH and Astrium UK for their support in preparing this paper.

\bibliographystyle{spphys}   
\bibliography{tinetti} %% sans espaces entre les fichiers !

\end{document}

%% file: scientific_objectives.tex
\begin{figure}[h]
\centering
\includegraphics[width=0.5\textwidth]{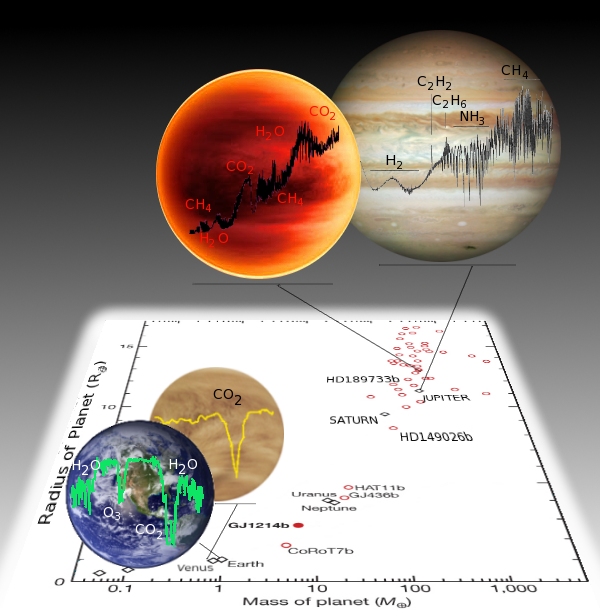} 
\caption{ Planets can be very similar in mass and radius and yet
be very different worlds, as demonstrated by these two pairs of examples.
A spectroscopic analysis of the atmospheres is needed to
reveal their physical and chemical identities.  } 
\label{obj}
\end{figure} 
%\vspace{0.3cm}
\section{Introduction}
The Exoplanet Characterisation Observatory, or EChO, is a proposed M class
mission currently under assessment by the European Space Agency (ESA) \footnote{http://sci.esa.int/echo}.
In this article we present scientific and technical information about the
proposed satellite.
\subsection{Scientific objectives}
%%%%%%%%%%%%%%%
The scientific objectives of EChO are to:
\begin{enumerate}
\item Measure the \textbf{\emph{atmospheric composition, temperature and 
albedo} } of  a highly
representative sample of known extrasolar planets, orbiting different stellar 
types (A, F, G, K and M). The sample will include hot, warm, and
habitable-zone exoplanets, down to the super-Earth size ($\sim 1.5$ Earth 
radii).  
The climate of a planet depends on the amount of stellar irradiation
reflected out to 
   space and absorbed.
 The combination of visible albedo and infrared temperature will be key to understanding how
the energy is redistributed. 
%%%%%%%%%%%%%%%%%%%%%%
\item Measure the spatial (vertical and horizontal) and temporal variability 
of the thermal/chemical atmospheric structure of hot giants, Neptunes and super-Earths orbiting bright stars.
The photometric accuracy of EChO at multiple wavelengths will be sufficient to
observe the planet not merely as day/night hemispheres or terminator but to divide the planet into longitudinal slices, hence producing coarse maps of  exoplanets (see \S 5.1 and 5.2).
Repeated ingress/egress measurements and phase light-curves for bright eclipsing hot
exoplanets will  advance atmospheric modelling efforts. This spatial/temporal differentiation is necessary to:
\begin{itemize}
\item Understand the relative importance of \textbf{\emph{thermochemical 
equilibrium, photochemistry}}, and transport-induced quenching in 
controlling the observed composition.  
\item 
Provide much needed constraints for \textbf{\emph{atmospheric 
dynamics and circulation models}}.  
 Longitudinal brightness maps obtained from the light curve of phase variations, observed by EChO, promise to be a powerful diagnostic tools for simulations of hot planets' atmospheric dynamics. \newline
 Vortices and waves are structures in exoplanet 
atmospheres that can produce observable temporal variability:
these are usually long-lived and evolve with characteristic periodicities \cite{Cho2003,Thrastarson2010}, which can be captured by EChO's observations (see \S 6.2). 
\end{itemize}
%%%%%%%%%%%%
\item Investigate the complex \textbf{\emph{planet-star interaction}}. 
Proper characterisation of a planet's host star  is  key  to the 
interpretation and to the
understanding of planetary data. 
Monitoring stellar variability simultaneously with acquiring the
data, from which the exoplanet atmosphere will be measured, is a key
aspect of the EChO mission.  
\item Constrain the models of \textbf{\emph{internal structure}}. 
EChO will  be able to measure with exquisite accuracy the 
depth of the primary transit and thus the planetary size, but the major 
improvements for interior models will come from its ability to fully 
characterise the atmosphere in its composition, dynamics and structure. 
\item Improve our understanding of \textbf{\emph{planetary formation/evolution}}
mechanisms.  High resolution spectroscopy will provide important information 
about
the chemical constituents of planetary atmospheres, and  this is expected to 
be related to both the formation location,
and the chemical state of the protoplanetary disk. 
\item Explore the thermal/chemical variability along the orbit of  \textbf{\emph{non transiting
exoplanets}, especially in high-eccentric orbits. }  This work was pioneered by Harrington \emph{et al.} \cite{Harrington2006,Crossfield2010} who made
phase curve measurements of the non-transiting exoplanet Ups And b. 
In contrast to the nearly circular orbits of the planets in the Solar System, highly eccentric orbits ($e \ge 0.3$) are common among the exoplanets discovered to date (e.g. HD80606b \cite{Laughlin2009}).
While non-transiting planets will not be a primary goal, EChO will give us the unique opportunity of studying the chemistry and 
thermal properties of very exotic objects.  
\end{enumerate}
%%%%%%%%%%%%%%%%%%%%
%\vspace{0.3cm}
 On top of that, EChO could: 
\begin{itemize}
\item  \textbf{\emph{ Search for Exomoons.} } We estimate that moons 
down to
0.33$R_{\oplus}$ would be detectable with EChO for our
target stars. Whilst \textit{Kepler} may also be able to detect
exomoons  \cite{Kipping2010}, EChO can obtain NIR light curves which exhibit highly reduced
distortion from limb darkening and stellar activity e.g. spots.
Additionally, multi-colour light curves significantly attenuate
degeneracy of fitted limb darkening parameters across all wavelengths (see \S 6.5).
\item Identify \textbf{\emph{potential biosignatures}} in the atmospheres of
super-Earths in the habitable zones of late type
stars.  
The study of super-Earths in the habitable zone of stars cooler than the Sun
will challenge the paradigm of the Earth-twin orbiting a Sun-twin as the only 
possible cradle for life\cite{Segura2005,Grenfell2011a}.
\end{itemize}
The science return of EChO is summarised in \cite{TinettiIAU}.
%%%%%%%%%%%%%%
\begin{figure}[h]
\centering
\includegraphics[width=0.5\textwidth,trim=0cm 0cm 5cm 2cm, clip=true]{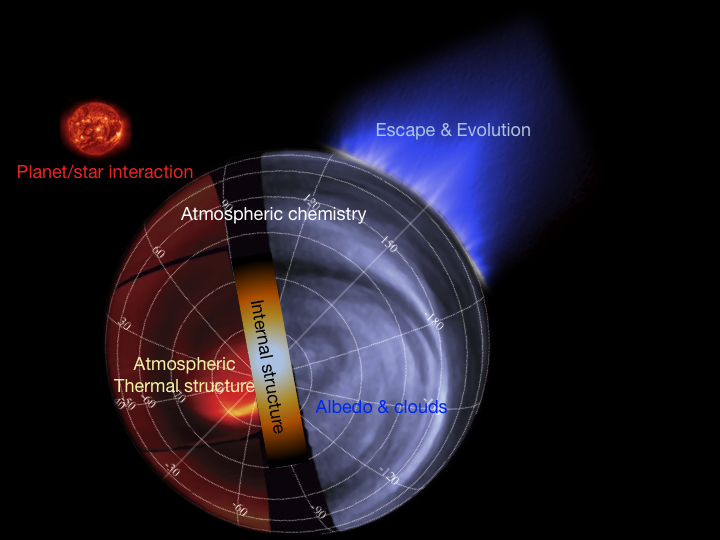} 
\caption{ Scientific objectives of EChO.  } 
\label{obj1}
\end{figure} 
%%%%%%%%%%%%%%%%%%
\subsection[Targets observed by EChO]{Targets observed by EChO: which, how many, why?}  
Table \ref{targets} lists the combinations of exoplanets/parent stars 
which will be observed by EChO. The planetary classification is done in terms of size 
and temperature: these
 two parameters combined with the stellar type, give the  star-planet contrast at  different wavelengths. 
The contrast, multiplied by the stellar luminosity, determines the 
feasibility and the integration time of the eclipse observations. For primary transit observations,  the  parameter to consider is the atmospheric scale height:
the hotter is the atmosphere and the lighter is the main atmospheric component and the planetary mass, the easier is the primary transit observation. While EChO will be able to observe the 
secondary eclipse for all types of planets listed in Table \ref{targets}, primary transit observations will be guaranteed only for  exoplanets with a light 
main atmospheric component and/or relatively high temperature, nominally gas-giants, Neptunes and a sub-sample of super-Earths  with those characteristics  (see \S 5.4).

\begin{table}[h]
%\rowcolors{3}{red!10}{}
\centering  
\begin{tabular}{|l|c|c|c|}
\hline 
 \qquad \qquad \qquad \, \, \textbf{Size}  &
 Jupiters &
 Neptunes &
Super-\\
 \textbf{Temperature} &
 &
 &
Earths \\
\hline
\rowcolor{orange!30}{Hot {\textgreater} 700 K}&{F,G,K,M }&{G,K,M }&{M}   \\\hline
 %%%%%%%%%%%
 \rowcolor{yellow!20}{Warm: 400-700~K}&{F,G,K,M }&{G,K,M }&{M}       \\\hline
  \rowcolor{green!20}{Temperate: 250-350~K}&{F,G,K,M }&{G,K,M }&{late M}       \\\hline
\end{tabular} 
\caption{Type of planets and corresponding type of star observable by EChO. See \S 5.5 and 9 for additional information.}
\label{targets}
\end{table}

 \emph{Super-Earths will be given high priority}. In practice, we expect that the atmospheres of terrestrial planets will show great diversity, well beyond the limited number of cases found in our
  Solar System (Earth, Mars, Venus, Titan).
On top of that, EChO will  easily observe a large number of gas giants and 
Neptunes orbiting different types of stars, with a variety of masses, radii and temperatures. A  significant subset of those will be observed  at high spectral  resolution and with 
multiple visits,  to monitor spatially and temporally resolved patterns due to photochemistry and dynamics.

Today, EChO could observe $\sim$80 known  gas and icy giants transiting  stars brighter than   V$\sim$12 mag \cite{schneider}, four transiting super-Earths -- the hot CoRot-7b
\cite{Leger2009}), Kepler-10b \cite{Batalha2011}, 55Cnc-e \cite{Winn2011}, the warm GJ~1214b \cite{Charbonneau2010} and HD 97658b \cite{Henry2011}--, a non-transiting hot gas-giant, Ups And b \cite{Marcy1996}) and a non-transiting hot super-Earth, GJ 876d \cite{Rivera2005}.  We note that GJ~581 is a nearby star with at least 3 super-Earths orbiting in the
vicinity of its habitable zone \cite{Mayor2009}. These super-Earths would be ideal  for EChO if they transited.
Although today the available target sample is still biased towards more massive planets, 
 HARPS data show that more than 40\% of stars have planets with 
masses below 50 Earth-masses
and  30\% of stars with planets below 30 Earth-masses \cite{Mayor2009,Mayor2011}. 
From preliminary analysis of Kepler, the occurrence of 2-4 $R_{\oplus}$ planets in the Kepler field linearly increases with decreasing stellar temperature, making these small planets seven times more abundant around cool stars (3600-4100~K) than the hottest stars in our sample (6600-7100~K) \cite{Borucki2011,Howard2011}.
Statistical estimates  from microlensing surveys \cite{Cassan2011}  indicate that 
on average, every star in the milky way has one or more planet at least 5 times the mass of the Earth in the orbit range 0.5-10 AU.
See \S 4 for further discussion on the targets available in 2020.

%%%%%%%%%%%%
\begin{figure}[h]
\includegraphics[width=0.5\textwidth,trim=0.5cm 3cm 1.5cm 0cm, clip=true]{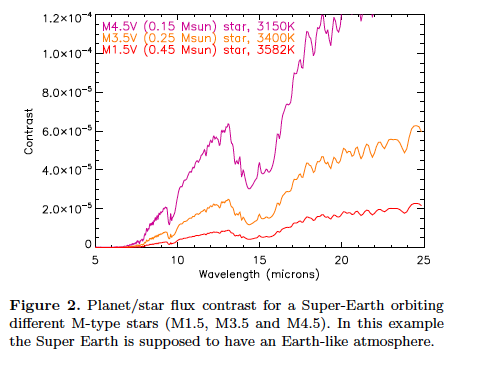} 
\caption{ Good reasons for  considering super-Earths around M-dwarfs: the cooler the star and the smaller its radius, the better the contrast star-planet, as this simulation shows \cite{Tessenyi2011}.  }
\label{mastar1}
\end{figure} 
%%%%%%%%%%%%%%%

The most favourable star-planet contrast  in the case of the smallest targets, 
is obtained by observing planets around stars smaller and colder than our 
Sun, typically M-dwarfs.
There are several advantages in selecting this star-planet combination.
Cool stars of spectral type M comprise about  75\% of all stars,
both in the Solar neighbourhood and in the Milky Way as a whole. At the time of writing a new catalog for bright M-dwarfs has been published by Lepine and Gaidos \cite{Lepine2011}, presenting a new sample of 8889 M dwarfs with J$\le$10. M-dwarfs range in mass from about 0.5 M$_{\odot}$ to less than 0.1 M$_{\odot}$,
with associated reductions in heat and brightness. The sheer abundance
of M dwarfs throughout our Galaxy ensures that a large fraction of
exoplanetary systems will be centred on red suns. To date, radial
velocity searches have detected $\sim$20 planetary systems around M stars. As a 
group,
these systems harbour lower-mass planets orbiting at smaller semi-major
axes than those around Sun-like stars, supporting the assumption that
system architecture scales roughly with stellar mass and thus with
spectral type. 
%%%%%%%%%%%%%%

Given their meagre energy output, the habitable zones of M dwarfs, like
their ice lines, are located much closer to the primary than those of
more massive stars (Fig.~\ref{mastar}). While 0.10 AU and 0.19 AU are reasonable numbers
for the inner and outer boundaries of the habitable zones of larger M
dwarfs, with masses of about 0.4 M$_{\odot}$ (e.g., GJ 436), the corresponding
boundaries shrink to about 0.024-0.045 AU for the smallest members of
the class, with masses of about 0.1 M$_{\odot}$ \cite{Mandell2007}. 
This works in EChO's
favour, as the short orbital period (ranging from one week to one month depending on the M type)  will allow the observation of several tens (or even hundreds for late M) of transits during the life-time of the mission.
 The observation of the atmosphere of a terrestrial planet in the habitable zone of F, G, K type of star would, by contrast, be impractical with the transit technique. For instance one could only afford to observe $\sim5$ transits 
  in the lifetime of the mission for a terrestrial planet in the habitable zone of a Sun type star (one transit every calendar year!), which is too little time to retrieve a useful spectrum with appropriate S/N.

%% file: observational_strategy.tex
\section{Observational  strategy and requirements}
\label{strategy}
\subsection{Observational techniques used by EChO}
EChO will probe the atmospheres of extrasolar planets combining three 
techniques, making use of a) planet transits, b) secondary eclipses, 
and c) planet phase-variations, which  will also be used for 
non-transiting planets. In all cases, instead of spatially separating the 
light of the planet from that of the star, EChO will use temporal variations
to extract the planet signal.
\begin{figure}[h]
\begin{center}
\includegraphics[width=0.5\textwidth,trim=0cm 0cm 0cm 1cm, clip=true]{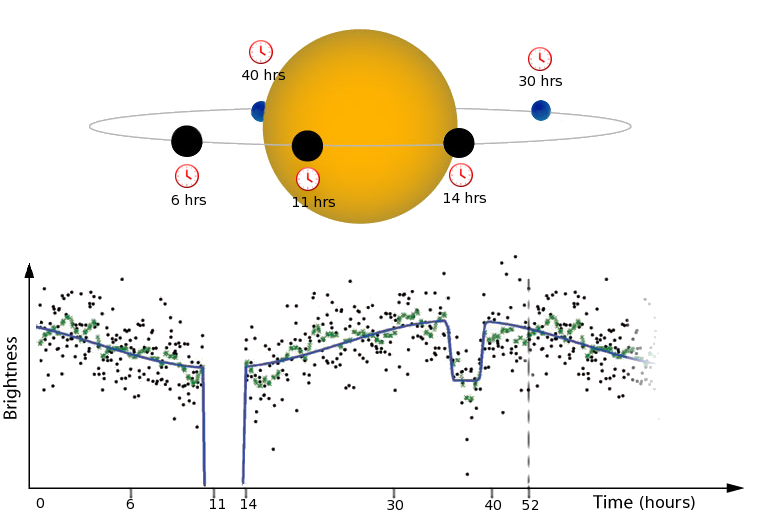}  
\end{center}
\caption{Optical phase curve of the planet HAT-P-7b observed by Kepler \cite{Borucki2009} showing primary transit and secondary eclipse measurements. }
\label{map}
\end{figure}

\paragraph{Transmission spectroscopy}
When a planet partially eclipses its host star, star-light filters through 
the planet's atmosphere, adopting a spectral imprint of the atmospheric 
constituents. By comparison of in-transit with out-of-transit observations,
this planet absorption is distilled from the absorption spectrum of 
the host star \cite{Seager2000,brown,tinettia}. Transmission spectroscopy probes the 
high-altitude atmosphere at the day/night terminator region 
of the planet. Typically, absorption features scale with the atmospheric 
scale-height, which mainly depends on the temperature and mean molecular 
weight of the atmosphere. The first successes of exoplanet transmission 
spectroscopy were in the UV and visible \cite{charbonneau2,VidalMadjar2003,Knutson2007,Pont2008,redfeild,snellen,bean2010}, and have  been later extended to the near- 
and mid-infrared \cite{Knutson2007a,tinettib,tinettic,swain2008a,beaulieu,beaulieu2010a,Beaulieu2010b,Agol2010}.

\paragraph{Secondary eclipse spectroscopy}
When a planet moves behind its host star (the secondary eclipse), the planet 
is temporarily blocked from our view, and the difference between in-eclipse 
and out-of-eclipse observations provides the planet's dayside spectrum. 
In the near- and mid-infrared, the radiation is dominated by thermal emission,
modulated by molecular features \cite{deming,charbonneau3,swain2009a,swain2009b,swain2010,Stevenson2010}. This is highly 
dependent on the vertical 
temperature structure of the atmosphere, and probes the atmosphere at much 
higher pressure-levels than transmission spectroscopy. 
At visible wavelengths, the planet's spectrum is 
dominated by Rayleigh and/or Mie scattering of stellar radiation \cite{Rowe2006,Kipping2011}. For the 
latter, clouds can play an important role.
%\begin{figure}[t]
%\includegraphics[width=0.9\linewidth]{Figures_originals/xo1b.jpg} 
%\includegraphics[width=0.8\linewidth]{Figures_originals/snellen.jpg} 
%\caption{Atmospheric signatures of the planet XO-1b obtained with 
%primary transit (Hubble) \cite{tinettic} and phase-curve of the planet CoRot-1b  
%\cite{Snellen2009}. }
%\end{figure}

\paragraph{Planet phase-variations}
In addition, during a planet's orbit, varying parts
of the planet's day- and night-side are seen. By measuring the minute 
changes in
brightness as a  function of orbital phase, the longitudinal brightness
distribution of a planet can be determined. Since the typical time
scale of these phase-variations largely exceeds that of one observing
night and they are of very small amplitude, these observations can only be conducted from space. However, 
they can also be performed on non-transiting planets \cite{Crossfield2010}.
Phase-variations are important in understanding a planet's
atmospheric dynamics and the redistribution of absorbed stellar energy from 
their irradiated day-side to the night-side. 
Ground-breaking infrared 8~$\mu$m Spitzer observations of the
presumably phase-locked
exoplanet HD189733b have shown the night-side of this hot Jupiter to be
only about 300~K cooler than its day-side \cite{Knutson2007a}, implying
an efficient redistribution of the absorbed stellar energy. These
same observations show that the hottest (brightest) part of this planet
is significantly offset with respect to the sub-stellar point,
indicative of a longitudinal jet-stream transporting the absorbed heat
to the night-side. 
Towards the optical wavelength regime, an increasing contribution from
reflected light is expected (as with secondary eclipses), as is likely the 
case in CoRoT \cite{Snellen2009} and Kepler \cite{Borucki2009} light-curves. 
\begin{figure}[h]
 \centering\includegraphics[width=0.5\textwidth,trim=0cm 8cm 0cm 0cm, clip=true]{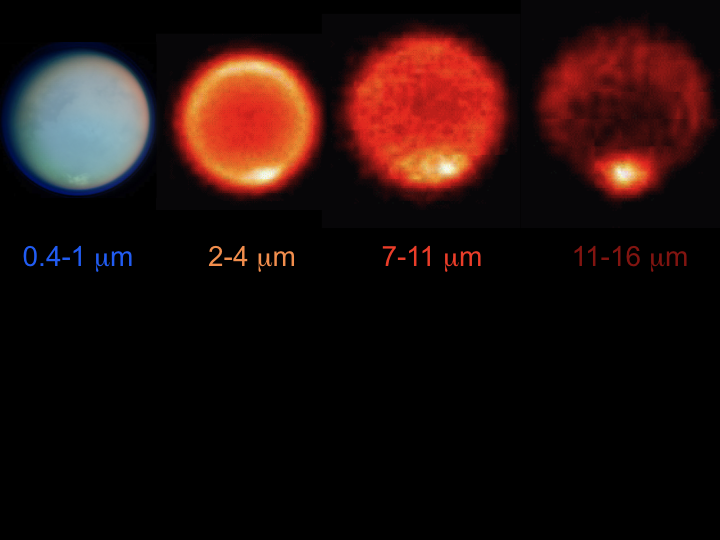}
\caption{Demostrator of possible results from exo-cartography of a  planet at multiple photometric bands. For hot giant planets orbiting a bright star, this can be achieved by EChO with $\sim$ 100 transits. For instance HD 189733b, can be mapped in 10 longitudinal slices with a spectral resolving power  R $\sim$20 in the IR and a S/N $\sim$ 100. By binning more spectral channels we can improve the spatial resolution.}
\label{map} \end{figure}

\paragraph{Spatial and temporal resolution}
During a primary transit we probe the planetary limb at the terminator,
whereas during secondary eclipse we probe the planetary hemisphere exposed to 
stellar radiation (day-side).  If we have transit, eclipse and phase-curve 
measurements, we can extract 
the spectrum of the un-illuminated (night-side) hemisphere \cite{KippingTinetti2010}.
Eclipses can be used as powerful tools to spatially resolve the photospheric 
emission
properties of astronomical objects. During ingress and egress, 
the partial occultation effectively
maps the photospheric emission region of the object being eclipsed \cite{Rauscher2007,Cowan2009}. Note that the different system geometries affect the orientation and shape of the eclipsing stellar limb and consequently the detailed
shape of the ingress/egress curves. Fig.~\ref{map} illustrates possible
results from exo-cartography experiments.
 The regime of atmospheric circulation present 
on hot, close-in exoplanets may be unlike any of
the familiar cases encountered in the Solar System. 
Key constraints 
will be placed by EChO on these models through repeated infrared measurements.
%%%%%%%%%
\subsection{Interpreting exoplanet spectra}
 Key species which should be observable by EChO are given in Table \ref{molecules}.
Their complexity, together with the potential for overlapping 
molecular bands,
means that the spectra can only be interpreted by comparing them to 
detailed atmospheric
models.

In an emission spectrum, measured through secondary eclipse observations in the IR, 
molecular signatures can appear either in absorption, emission,
or both, depending on the shape of the pressure-temperature profile and the 
molecular vertical
mixing ratio \cite{Goody1989,Hanel}.
 Spectral 
retrieval
methods and forward models are used to infer the presence and abundance of 
specific molecules
and, in the case of an emission spectrum, the pressure-temperature profile; 
this can lead to a
natural ambiguity between composition and 
temperature \cite{swain2009b,madhu,TinettiFaraday,Lee2011,Line2011}.
Once the composition and temperature structure has been determined, knowledge 
of the
atmospheric chemistry is inferred from the abundance estimates and vertical 
mixing ratios of
individual molecules. For example, if the mixing ratio of CO$_2$ is higher than 
would be expected
from purely equilibrium chemistry, a non-equilibrium chemistry mechanism (such 
as photochemistry) may be needed to explain the additional CO$_2$.
%\begin{figure}[h]
%\includegraphics[width=\linewidth]{Figures_originals/proposalNov2-img16.jpg}
%\caption{Emission photometry and spectroscopy the hot-Jupiter HD 209458 b \cite{swain2009b}.  The near-infrared and mid-infrared 
% observations are compared to synthetic spectra for four models that illustrate 
% the range of temperature/composition possibilities consistent with the data. 
% For each model case, the molecular abundance of CH$_4$, H$_2$O, \& CO$_2$ and the location of the tropopause is given, these serve to illustrate how the 
% combination of molecular opacities and the temperature structure cause significant departures from a purely single-temperature thermal emission 
% spectrum. Note that the current mid-infrared data shown here are not contemporaneous with the near-infrared data, and attempting to ``connect" these data sets with a model 
% spectrum is potentially problematic if significant variability is present. This will not be an issue for EChO.}
% \label{specretr} \end{figure}
\subsection{Justification of  wavelength coverage and spectral resolution} 
For secondary eclipse measurements in the thermal regime (emission 
spectroscopy), retrieving abundances will require the
simultaneous retrieval of the thermal profile, i.e. distinguishing
between which features are
in emission and which ones are in absorption. This will be made easier
when bands of different intensities are used for a given molecule.
EChO will obtain broad, instantaneous and simultaneous
spectral coverage from  the visible 
(0.4\,$\mu$m) to the mid-infrared (16\,$\mu$m). Broad wavelength coverage 
enables
resolving the temperature/~composition ambiguity in an emission spectrum; 
simultaneous measurement of VIS-IR wavelengths allows planetary and stellar 
variability to be characterised and understood. 
\begin{table}[h!]
%\begin{minipage}{0.5\linewidth}
\footnotesize
\rowcolors{1}{green!20}{}
%\begin{tabularx}{\linewidth}{|X|X|X|X|X|}
\begin{tabular}{|p{1.cm}|p{1.4cm}|p{1.4cm}|p{1.4cm}|p{1.4cm}|}
\hline
 	& 0.4-1\,$\mu$m	& 1-5\,$\mu$m	& 5-11\,$\mu$m 	& 11-16\,$\mu$m \\
\hline
 \emph{R, baseline} 	&  $\sim$Few tens	&   300 &  $\ge$30	&  20  \\
 \hline
\emph{ R, desired }	&  300	&  300  &  300 	&  300  \\
\hline
{*}H$_{2}$O	& 0.51, 0.57, 0.65, 0.72, 0.82, 0.94 & 1.13, 1.38, 1.9, \textbf{2.69} & 6.2 &  conti-nuum  \\ 
{*}CO$_{2}$  &   - 	& 1.21, 1.57, 1.6, 2.03, \textbf{4.25}	& -	& \textbf{15.0} \\  
C$_{2}$H$_{2}$& -	& 1.52, \textbf{3.0}		& 7.53	& \textbf{13.7}  \\  
HCN	 & -	& \textbf{3.0}			& -	& \textbf{14.0} \\  
C$_{2}$H$_{6}$ & -  & 3.4				& -	& \textbf{12.1} \\   
O$_{3}$	&  0.45-0.75 (the Chappuis band)	& 4.7				&  9.1, \textbf{9.6} 	& 14.3  \\
HDO	& -	& 2.7,3.67			& 7.13	& - \\
{*}CO	& -	& 1.57, 2.35, \textbf{4.7}	& -	& - \\
O$_{2}$	&  0.58, 0.69, 0.76, 1.27	& 	-			&  -	&   -  \\
NH$_{3}$ 	& 0.55, 0.65, 0.93	& 1.5, 2, 2.25, 2.9, \textbf{3.0}	& 6.1, \textbf{10.5} & - \\
PH$_{3}$  & - 	& 4.3				& 8.9, 10.1 & - \\
{*}CH$_{4}$	& 0.48,  0.57. 0.6,  0.7, 0.79,  0.86,  & 1.65, 2.2, 2.31, 2.37, \textbf{3.3} & 6.5, \textbf{7.7} & - \\
CH$_{3}$D	& ?	& 3.34, \textbf{4.5}		& 6.8, 7.7, \textbf{8.6} & - \\
C$_{2}$H$_{4}$ & - 	& \textbf{3.22}, 3.34		& 6.9, \textbf{10.5} & - \\
H$_{2}$S	& -	&     		2.5, 3.8 ...		&  7	& - \\
SO$_{2}$	& -	& 		4		&  \textbf{7.3}, 8.8	& - \\
N$_{2}$O	& -	& 		2.8, 3.9, \textbf{4.5}		& 7.7, 8.5 	& - \\
NO$_{2}$	& -	& 		3.4		& \textbf{6.2}, 7.7	&  13.5 \\
H$_{2}$	& -	& 2.12				& -	& - \\
H$_{3}^{+}$ & -	& 2.0, 3-4.5			& -	& - \\
He	& -	& 1.083				& -	& - \\
{*}Na 	& 0.589	&     1.2				& -	& - \\
{*}K	 & 0.76   	& -				& -	& - \\
TiO	&  0.4-1	&    1-3.5			& -	& - \\
VO	&  0.4-1	&    1-2.5			& -	& - \\
FeH	&  0.6-1	&   1-2				& -	& - \\
TiH	&  0.4-1	&   1-1.6				& -	& - \\
%CrH	& 	& -				& -	& - \\
% Rb, Cs &  yes	& -				& -	& - \\
Rayleigh & 0.4-1 & -		& -	& - \\
Cloud/ haze &   yes	& possible	& silicates, etc. & - \\
H  H$\alpha$ & \textbf{0.66} & &  &  \\
H  H$\beta$ & 0.486 & &  &  \\   
% LiI  &  \textbf{0.67}	& -				& -	& - \\
 Ca &     0.8498,  0.8542,   0.8662      &			& -	& - \\
\hline
\end{tabular}
%\end{minipage}
\caption{Main spectral features between 0.4 and 16\,$\mu$m. The asterisk indicates  the 
molecular/atomic species already detected in the atmospheres of exoplanets.
At wavelengths shorter than 2\,$\mu$m spectroscopic data are often not complete, 
so that the use of this region is much more difficult for band identification 
and analysis. The main bands are illustrated in bold.}
\label{molecules}
\end{table}
% \begin{figure}[h]
% \begin{minipage}{0.7\linewidth}
% \includegraphics[width=\linewidth,trim=0cm 0cm 0.8cm 0cm,clip=true]{Figures_originals/earth-venus-mars-spectrum.jpg}
% \includegraphics[width=\linewidth]{Figures_originals/saturn_jupiter.jpg}
% \end{minipage}
% \begin{minipage}{0.27\linewidth}
% \caption{The 11-16\,$\mu$m band is a key band for retrieving the thermal profile in planetary atmospheres (especially exoplanets in the habitable zone).  \newline
% From a chemical point of view, the main loss would be the CO$_2$ band at 15\,$\mu$m for terrestrial planets and the Q-branches of hydrocarbon species for hydrogen-rich atmospheres. }
% \end{minipage}
% \label{mirchannel}
% \end{figure}
Having signatures in both the
reflected and thermal regions will greatly help the abundance and
temperature structure retrieval. Moreover, monitoring stellar variability 
simultaneously with the acquisition of 
data from which the exoplanet atmosphere will be measured is a key 
aspect of the EChO mission: the light 
variations caused by magnetic activity can hamper the extraction of the 
exoplanet atmosphere signal and a need arises to diagnose stellar variability 
mostly in the 
near-IR and mid-IR continuum. Such variations are associated with active 
regions (star spots and bright spots or faculae) coming on and off view 
as the star rotates and also from intrinsic variability of such active 
regions. Both variations can  occur on 
relatively short timescales, comparable to those of the planet's orbital 
period, and thus impact directly on the combination of different epochs 
of eclipse data. The best available indicator of 
chromospheric flux in the wavelength ranges accessible to EChO is the 
hydrogen Balmer $\alpha$ line at 0.66\,$\mu$m. Emission in the core of the 
line appears as a consequence of chromospheric activity and thus can be 
used to monitor variations in the stellar chromosphere \cite{Walkowicz2009}. Observations in the visible range are thus essential to 
provide the stellar data needed for the measurement and interpretation 
of exoplanet atmospheres.
The ability to reach 0.4\,$\mu$m would be important for observing the contribution of Rayleigh scattering. For a cloud/haze free atmosphere this additional information 
is key to removing the degeneracy embedded in the measurements of the planetary radius at wavelengths where molecules absorb.

The resolving power given for the EChO base line will be sufficient not only to 
separate the bands but  go to the next step, i.e. detect the molecular 
features, retrieve
abundances, disentangle the contribution of different molecules if they overlap 
etc.
Table \ref{molecules} gives the most important molecular and atomic  
species likely to be present in planetary atmospheres and have a spectral 
signature in the wavelength region covered by EChO. 
We also indicate the spectral resolving power obtainable by EChO, in its baseline configuration, and the 
goal.

The 11-16\,$\mu$m band is crucial --particularly for the CO$_2$ band 
at 15\,$\mu$m--  for
retrieving the thermal profile in terrestrial atmospheres, 
especially planets in the habitable zone \cite{Hanel}. 
A resolving power of 
R =300 in that channel instead of the currently proposed 20, would make
it possible to 
separate 
C$_2$H$_2$ from HCN and greatly improve the temperature structure 
retrieval for gas-giants and Neptunes.
\subsection{Justification of signal-to-noise, timing, calibration, observational strategy} 
\label{SNTC}
\begin{figure}[h]
\centering
\includegraphics[width=0.5\textwidth]{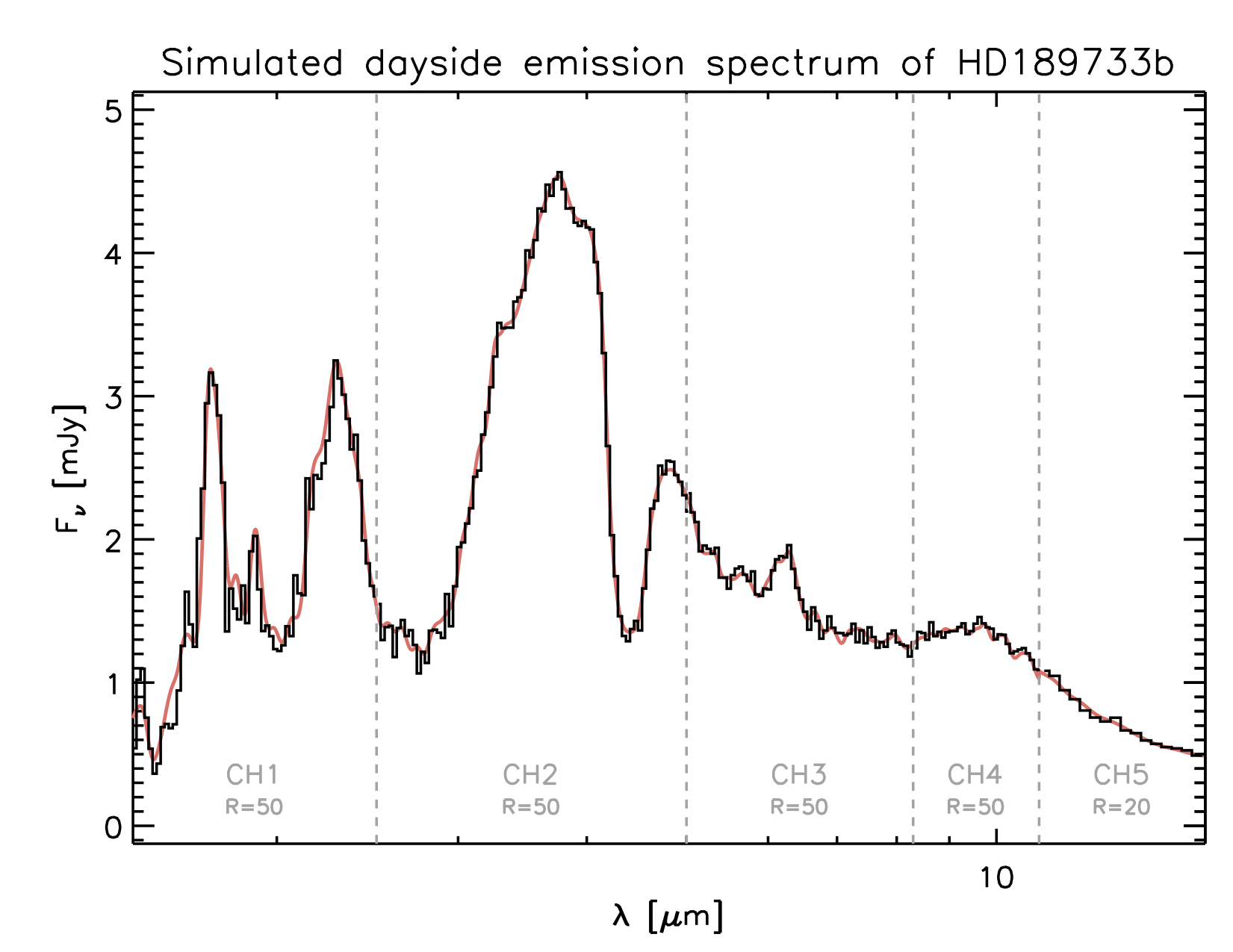}  
\includegraphics[width=0.5\textwidth]{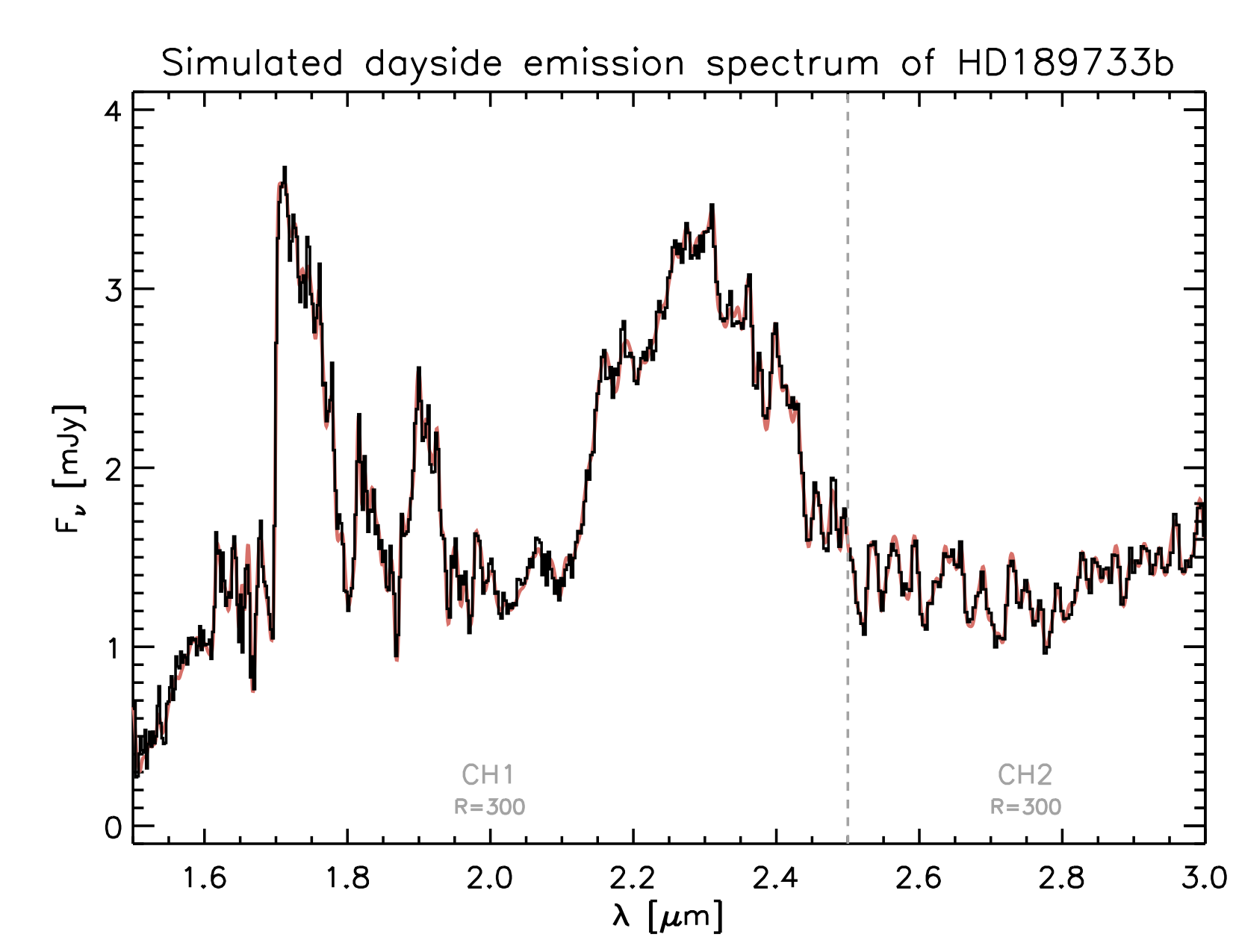}
\caption{Simulations of EChO observations of  the hot-Jupiter HD189733b. Top: dayside emission spectrum  at resolution R=50, single eclipse. 
Bottom: NIR zoom of dayside emission spectrum  at resolution R=300, averaged over 50 eclipses. Total observing time 8 days, which could be done over 3.5 months.}
\label{roy}
\end{figure}

\begin{figure}[h!]
\centering
\includegraphics[width=0.45\textwidth]{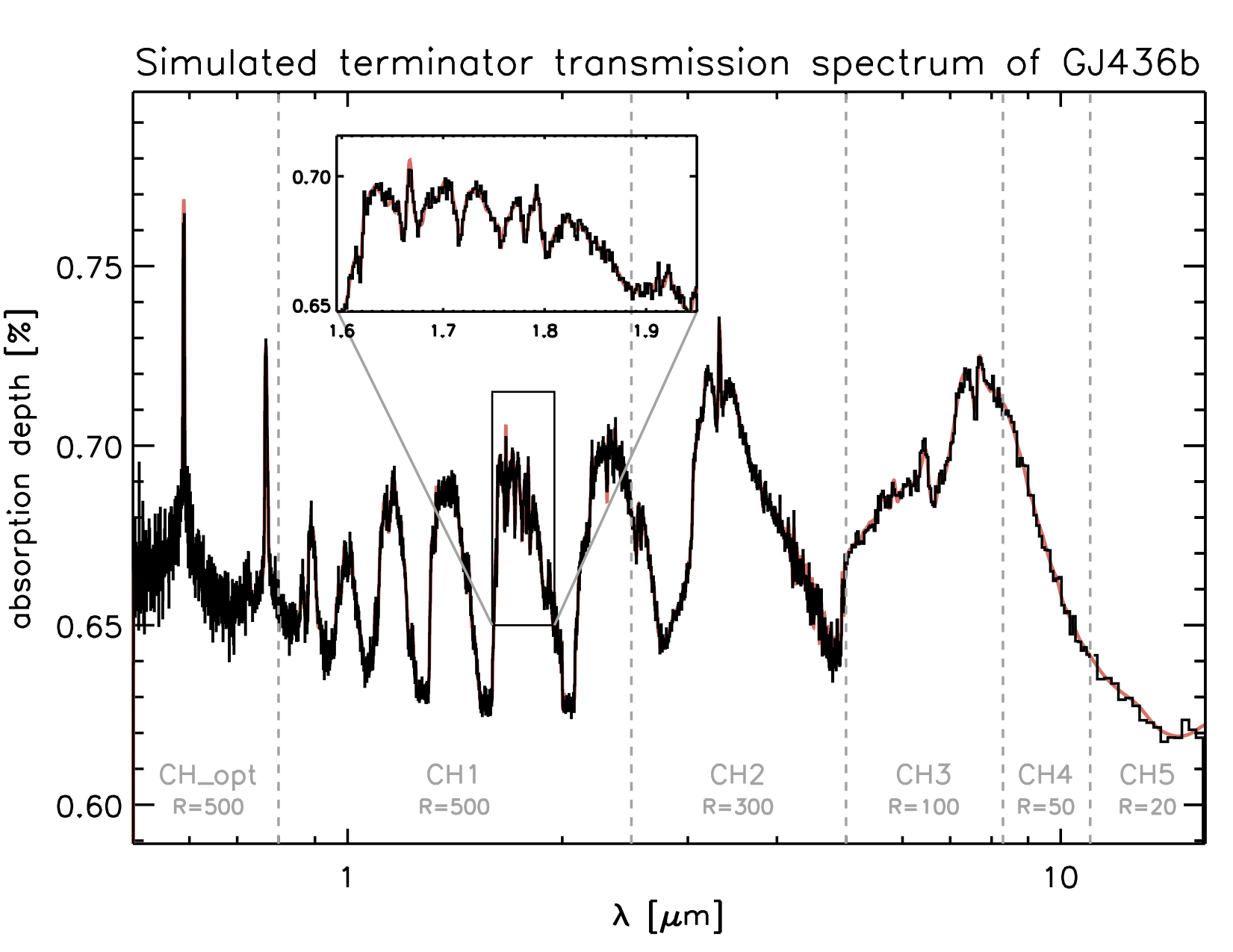}  
\includegraphics[width=0.45\textwidth]{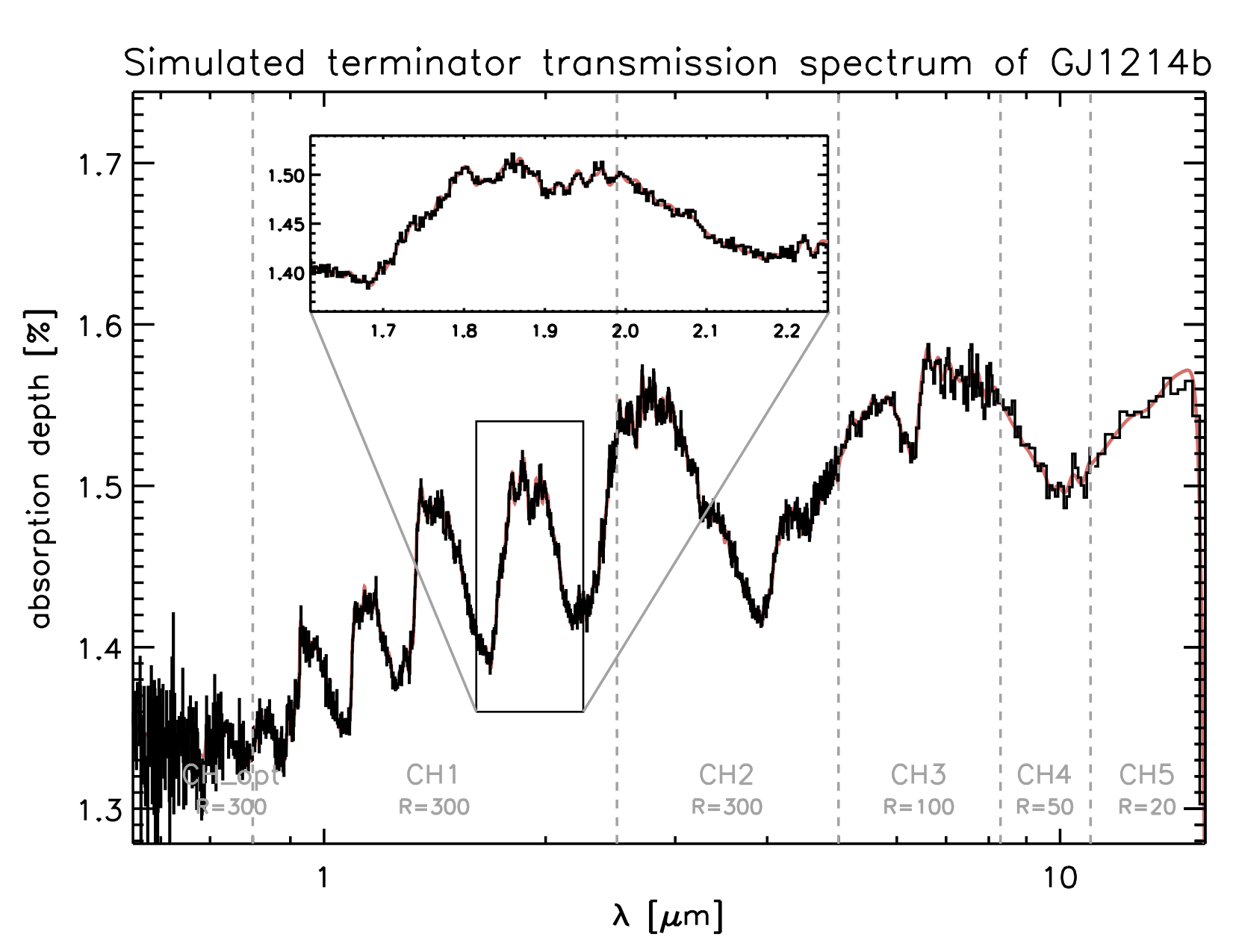} 
\includegraphics[width=0.35\textwidth]{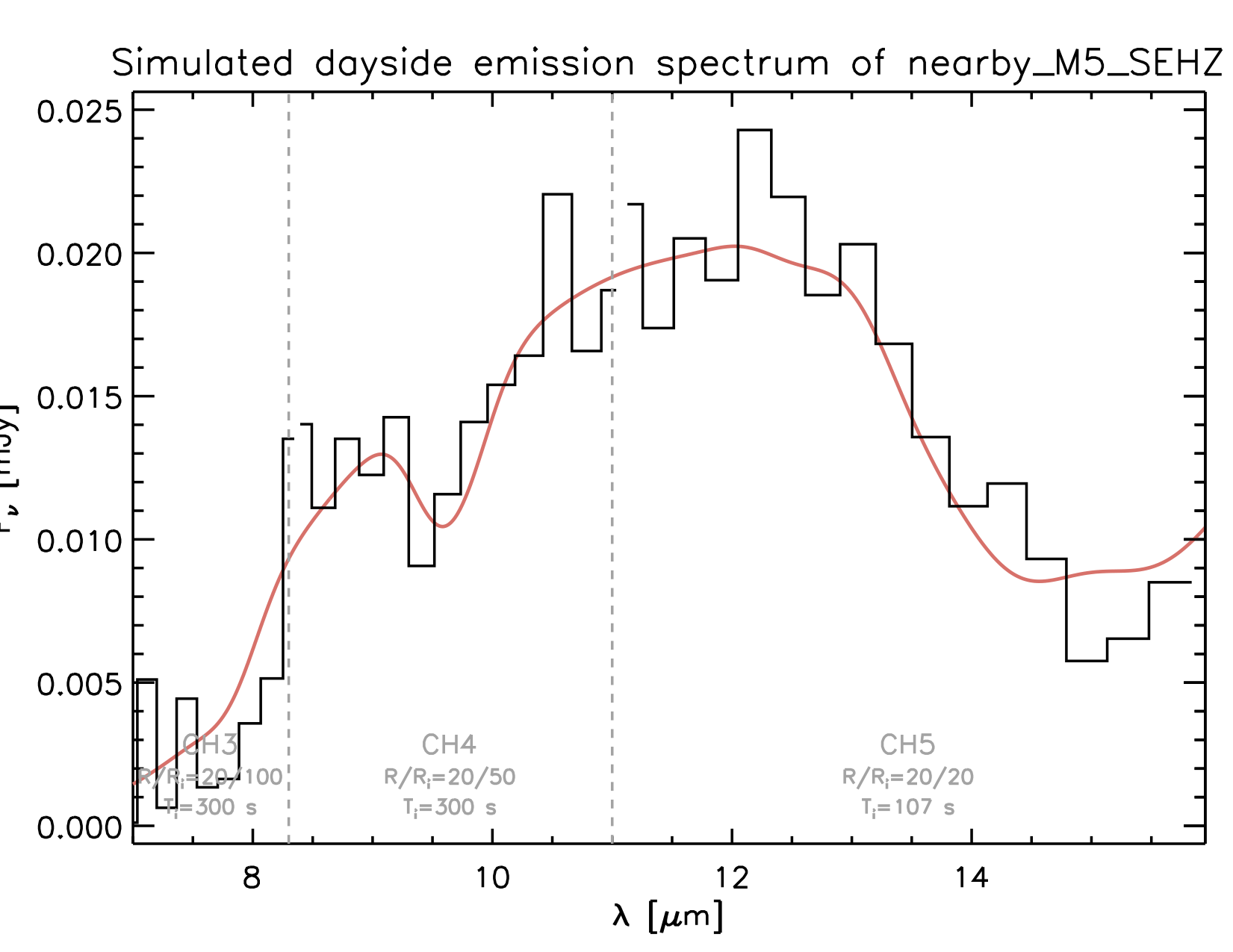}
\caption{ Simulations of EChO's performances.   1)   Transmission spectrum of warm-Neptune GJ 436b (top) at intrinsic instrumental resolution, averaged over 50 eclipses. 
Total observing time 6 days, can be done over 4 months.   2)  High resolution (R=300) transmission spectrum of warm super-Earth GJ1214b, averaged over 300 transits. Total observing time 3 weeks which could be done over 1.3 years (center).
3) Emission spectrum  of a favourable case of a temperate super-Earth  with an Earth-like atmosphere orbiting a bright late M.
(bottom).  }
\label{roy1}
\end{figure}

The integration time needed to observe specific targets is
based on the time length required to obtain spectra of transiting exoplanet atmospheres, given a defined spectral resolution and signal-to-noise ratio. The estimated time is based
on the contrast ratio of the flux from the planet over the flux
from the star in a selected wavelength region and on the instrument parameters. EChO's characteristics were planned to guarantee the required performances.
The fluxes
were obtained from synthetic spectra and blackbody curves. We show here a few  key examples (Tables    \ref{tab:jup},  \ref{tab:hse}, \ref{tab:hzse}, Fig. \ref{roy}, \ref{roy1}).

According to our simulations, for spectroscopic observations (i.e. R$\ge$ 10) we need stellar targets brighter than V$\sim$12 for F, G and K stars, and brighter than K$\sim$9 for M dwarfs (see Tables \ref{tab:jup},  \ref{tab:hse} and \ref{tab:hzse}).
For giants orbiting  fainter stellar companions (V $ \le$ 15) such as typical CoRoT and Kepler targets, EChO will be able to  observe them in 2 or 3 large photometric bands (VIS, NIR and MIR).
There are currently $\sim$80 known transiting sources which are within the sensitivity range of EChO's spectroscopic capabilities (see ESA EChO - Science Requirements Document, http://sci.esa.int/echo).
\begin{table}[h]
\centering
   \rowcolors{4}{red!20}{}
   \begin{tabular}{|l|c|c|c|ccccc| }
   \multicolumn{9}{l}{Secondary eclipse, R=300, SNR=50, averaged over 5-16\,$\mu$m } \\
\hline
Star & T & R &contr.&\multicolumn{5}{ c|}{Magnitudes in V band} \\
type & (K)&($R_{\odot}$) &$10^{-3}$&  5 & 6 & 7 & 8 & 9 \\
\hline
F3V&6740&1.56&1&7&18&51&156&$<$R
\\G2&5800&1&2.9&0.7&1.8&4.7&14&45
\\K1&4980&0.8&5.6&0.2&0.4&1&2.9&9\\
\hline
\multicolumn{9}{l}{Primary eclipse, R=100, SNR=50, averaged over 5-16\,$\mu$m } \\
\hline
F3V&  6740 & 1.56 & 0.28 & 32 & 82 & 213 & $<$R & $<$R \\
 G2V& 5800 & 1 & 0.68 & 4 & 10 & 26 & 70 & 198 \\
K1V&  4980 & 0.8 & 1 & 1.6 & 4 & 10 & 26 & 72 \\
\hline
\end{tabular} 
   \caption{Integration times, in numbers of transits, for a hot-Jupiter in primary transit  (lower) and  in secondary eclipse (upper).}
   \label{tab:jup}
\end{table}
\begin{table}[h]
\centering
   \rowcolors{3}{orange!20}{}
   \begin{tabular}{|c|c|c|c|ccccc|}
\hline
M & T & R & contr.  &\multicolumn{5}{c|}{Magnitudes in K band} \\
 type & (K)&($R_{\odot}$) & $10^{-4}$ &  5 & 6 & 7 & 8 & 9\\
\hline
M1.5V & 3582 &0.42 &1.4 &  14  & 36 & 95 & 258 & - \\
M3V & 3436  &0.30   &2.8 &  6 & 13 & 34 & 93 & 277 \\
M4V & 3230 &0.19    &7.7 &  1 & 2 & 6 & 18 & 52 \\
M5V & 3055 &0.15   &13.2 &  0.5 & 1 & 3 & 8 & 23 \\
\hline
\end{tabular} 
   \caption{Integration times, in numbers of transits, for a hot (850 K) super-Earth (1.6 $R_{earth}$) in secondary transit assuming a resolution of 40, SNR 
of 10 observations in the range 5-16$\mu$m } 
   \label{tab:hse}
\end{table}
%%%%%%%%%%%%%%%%%%%%%%
\begin{table}[h]
     \centering
 %    \tiny
     \rowcolors{3}{green!20}{}
%   \begin{tabular}{|l|c|c|c|c|ccccc|}
   \begin{tabular}{|p{0.5cm}|p{0.5cm}|p{0.5cm}|p{0.5cm}|p{0.5cm}|p{0.35cm}p{0.35cm}p{0.35cm}p{0.35cm}p{0.35cm}|}
 \hline
Star&T&R&P&contr.&\multicolumn{5}{c|}{Magnitudes in K}\\
type&(K)&($R_{\odot}$)&days&$10^{-5}$&5&6&7&8&9\\
\hline
M2V& 3522      & 0.38 &    30.6             & 0.9     & 72     &          &  & &  \\
     & 3475    & 0.34 &    26.6         & 1.2     & 45     &113&  &  &  
     \\M3V& 3436      & 0.30 &    23         & 1.5     & 32     & 81     &  &  & \\
     & 3380    & 0.25 &    19.3         & 2         & 20     & 52     &132&  &  
     \\M4V& 3230     & 0.19 &    12.7         & 4         &         & 18     & 46   &117& \\
     & 3150     & 0.17 &    10.7         & 5.2     &           & 12     & 32 & 80 &208
     \\M5V& 3055     & 0.15 &    8.7         & 6.9     &         &          & 19 & 49 &128\\
     & 2920     & 0.13 &    6.7             & 9.8     &         &          & 12 & 29 &76\\
\hline
\end{tabular}
\caption{Integration times, in numbers of transits, for a habitable-zone (320 K) super-Earth (1.6 $R_{\oplus}$) in secondary transit observed with a resolution of  10, SNR 5 and in the   $5-16\,\mu$m  wavelength range.} 
\label{tab:hzse}
   \label{tab:mainresults320}
\end{table}
\begin{table}[h]
\centering\begin{tabular}{|lc|lc|}\hline
HD 189733 &  & GJ 1214 &    \\ \hline
 albedo=0.05  & $\sim 4 \sigma$  &         albedo=0.3  & $\sim$0.3 $\sigma$          \\
 albedo=0.1  & $\sim 8 \sigma$ & albedo=0.3, hot  & $\sim$26 $\sigma$  \\
 albedo=0.2  & $\sim 15 \sigma$ & &   \\\hline
 \end{tabular}
 \caption{Measurement of planetary albedo with one eclipse in the optical for key examples of hot-Jupiters and super-Earths. The results strongly depend on the type of planet/star, distance to the star and albedo value. } 
 \end{table}

 We will perform high-precision in-flight calibration and monitoring of the responsivity of the observatory, in short the ``transfer function (TF)", by observing calibration stars. Our goal is to monitor the spectral shape of the TF as well as its absolute level to a precision better than a few times $10^{-5}$, such that uncertainties in the TF do not significantly affect the final quality of the  science spectra obtained. 
 The Kepler mission is generating an unprecedented set of lightcurves for stars, with the best precision and coverage ever achieved  \cite{Basri2011}.
 Detailed modeling of optical light curves measured by the Kepler satellite using stellar model atmospheres shows that the vast majority of G-dwarfs and selected A-dwarfs have an intrinsic stability in their infrared emissions of better than 3$\times 10^{-5}$ in overall flux and better than 10$^{-5}$ in shape, where the former offer the highest precision in absolute level and the latter in shape. While the actual stars used to calibrate EChO will not be those observed by Kepler, but rather a sample of nearby main sequence stars distributed over the sky, we have demonstrated the feasibility of reaching a calibration precision of order 10$^{-5}$ using stars. We will build up a calibration network and perform pre-flight characterisation of the potential calibrators using ground-based high-accuracy (relative) photometry in order to ensure sufficient stability of the stars in the network. 
 It is relevant to emphasize that all the timescales related to stellar activity patterns are very different
from the timescales associated to single transit observations (a few hours), and thus can be easily
removed.
\begin{figure}[h]
\centering
\includegraphics[width=0.8\linewidth]{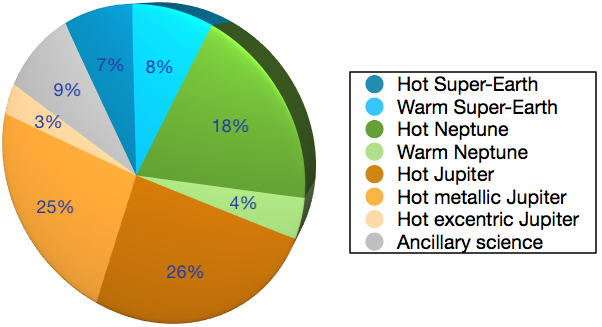}
\includegraphics[width=0.8\linewidth]{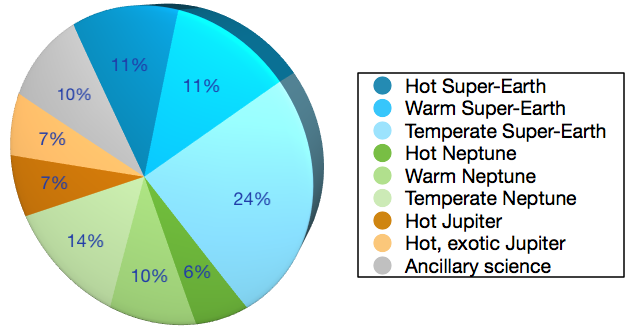} 
\caption{Top: partition of observing time on available sources for EChO today. Bottom: partition of observing time  for EChO in 2020. }
\label{source_partition}
\end{figure}
%\begin{figure*}[t]
%\centering\includegraphics[width=0.49\linewidth]{Figures_originals/source_partition.png}
%\centering\includegraphics[width=0.49\linewidth]{Figures_originals/source_partition_future.png}
%\caption{Left: example of observing time partition (by type of exoplanet) for an EChO mission observing a set of 52 sources already known today. Right: proposed observing time partition for an EChO mission based on sources known by 2020. }
%\label{source_partition}
%\end{figure*}
%\begin{figure}
%\centering\includegraphics[width=\linewidth]{Figures_originals/targets.png}  
%\caption{ Simulations of EChO observations for 50 known transiting  targets. }
%\end{figure}

%% file: synergy_othermissions.tex
\section[Synergies with other missions]{Differences and Synergies between EChO and future missions and facilities  }
In the upcoming decade, two important new facilities are planned to come on 
line  1) 
the space-borne James Webb Space Telescope (JWST) due to be launched in the
latter part of this decade, 
and 2) the next generation of extremely large telescopes, such as the European 
Extremely Large Telescope (E-ELT), with first light foreseen in 2018, or the Thirty Meter Telescope (TMT).   A significant advantage of EChO as a dedicated instrument is its ability to
provide the observations to fully test models: this requires observations of a large 
sample of objects, generally on long timescales, and cannot be efficiently pursued 
 with a multi-purpose facility such as JWST or the ELT.  Such a comprehensive approach 
holds out the possibility of discovering unexpected, ``Rosetta Stone"
objects, i.e.  objects that definitively confirm or disprove theories.

Compared to EChO, a 42\,m telescope such as the E-ELT has
two major advantages: a much larger collecting area and superior spatial resolution.   
Yet the E-ELT will suffer from the obvious limitations of every ground-based facility: namely much lower observing efficiency (e.g weather conditions, day/night cycles, observability of the target) and, more critically, more limited spectral coverage. A large fraction of the spectral range observed with EChO, in fact, is inaccessible from the ground (e.g. H$_2$O bands between 1-5\,$\mu$m, and the region between 5-8\,$\mu$m where there are key molecular lines).  Finally, long-term photometric stability, a key requirement for achieving the science goals of EChO, will never hardly be reached from the ground.

 EChO's telescope diameter might appear also small  compared to JWST. Yet to reach our science objectives, other parameters are  as critical: stability, spectral coverage, optimised detectors and
high degree of  visibility of the sky.
The key areas where EChO will excel compared to JWST are:
\begin{itemize}
\item	\emph{Dedicated mission}:  The main advantage of a dedicated mission such as EChO will be the design of an optimal scientific programme.  In the Design Reference Mission for JWST, at least 80-85\% of its time will be dedicated  to  non-exoplanet science.  This brings critical constraints on target observability and mission planning, especially for the planned EChO sample of time-critical observations of transiting exoplanets.  It will be impossible to perform large systematic surveys with many repeated observations of targets within the mission lifetime of JWST.  
\item \emph{Instantaneous wavelength coverage}. In contrast to JWST, EChO will simultaneously sample wavelengths from 0.4 to 16\,$\mu$m.  This is essential to study atmospheric variability and weather pattern.  JWST will need to observe at least four separate transit/secondary eclipse events to get similar, but still inferior, wavelength coverage.  
Important spectroscopic features like the CO$_2$ band at 15\,$\mu$m (see e.g. Figure 9), will have to be observed over more than one transit.  This has the potential to introduce fatal systematic errors in these  most sensitive measurements, especially for a planet orbiting an active star.  Conversely EChO will perform simultaneous observations over all wavelengths.
\item \emph{Long term photon-noise-limited stability of 10$^{-5}$}.  Although the thermal stability of JWST and its instruments will be very high, there are several factors which will limit its achievable precision: 1) The JWST instruments are optimised for background-limited observations and not for photon-limited observations, the latter case being appropriate for most of the nearest exciting targets; 2) The (generic) instruments on JWST contain many moving parts which will be a source of calibration uncertainty;  3) The segmented mirrors of JWST will exhibit low-level deformations over time causing temporal variations of the point spread function (PSF). EChO will not suffer from any of these problems. As long-term, high-level stability is essential for atmospheric variability and weather pattern studies, EChO will be superior in this respect.
\end{itemize}
EChO, specifically designed to reach $10^{-5}$ long-term, 
photon-noise-limited stability with detectors optimised for observing bright sources, will provide a critical scientific yield as a 
stand-alone observatory. However the synergy between EChO, JWST and the ELTs could be particularly powerful. 
EChO will guarantee a synoptic view over a wide variety of extrasolar planets by 
simultaneously measuring their emission/transmission spectrum from 0.4 to 16\,$\mu$m. A sub sample could be observed at higher resolution over a limited 
spectral window with JWST,
or very high resolution with an ELT which complement EChO observations. 

%% file: targets.tex
\section{Targets for EChO}
The main objective of EChO is to characterise spectroscopically the
atmospheres of exoplanets already discovered by other facilities at the
time EChO flies. 
To detect the chemical and thermal signatures of these remote worlds, the typical signal to aim at is between 10$^{-5}$ and 10$^{-3}$
 times the flux of the parent star: for this reason, the brighter the star in the wavelength range selected, the better.
Our simulations --validated against current observations of exoplanet atmospheres with Hubble, Spitzer and ground-based observatories-- indicate that the
brightness thresholds  for EChO's observations are: magnitude V$\sim$12 
for a G-K star, with an orbiting Jupiter or Neptune, and magnitude K$\sim$9  for ``Habitable-zone"
super-Earths around late M stars, see \S 5.4. 

Currently, about 80 of the $\sim 700$  identified exoplanets are transiting planets  with a stellar companion
satisfying those criteria \cite{schneider}. At present, the available sample for EChO includes few super-Earths and Neptunes, but it is still biased towards more massive 
planets. Thanks to ongoing and planned new surveys and facilities (the ESA-GAIA mission alone is expected to discover several thousands new exoplanets \cite{Casertano2008,Sozzetti2011}), a more complete reservoir of potential targets will become available  in the next decade.
The redundancy of observable targets will allow  the selection criteria for EChO targets to be refined. Below we give some approved projects and surveys that are likely provide
additional exciting targets for EChO over the next decade.

For solar-like stars, EChO
will observe stars up to a distance of 330 and 170 pc for a G0 and a K0, 
respectively, essentially the same volume explored by ground-based surveys. 
Ongoing 
and planned survey such as HATNet \cite{Bakos2002}, HAT-South \cite{Bakos2009}, WASP \cite{Pollacco2006},  LCOGT and XO \cite{XO} will significantly increase this number by expanding the parameter 
space (e.g. the surveyed spectral types, metallicity and sky coverage).

Recently several ground-based surveys devoted to the detection of planets around M stars have been launched.
 In particular programmes 
aiming to detect habitable-zone super-Earths around M-dwarf stars have started
and 
others are planned. All these projects will deliver good 
targets for EChO in addition to those already known such as the Neptune-like GJ 436b \cite{Gillon2007} and the super-Earth, GJ1214b \cite{Charbonneau2010}. 
These surveys include:
\begin{itemize} 
\item MEarth \cite{Nutzman2008} aims to
monitor  late M-dwarfs (R $\le$ 0.33 R$_{\odot}$) 
taken from the Lepine catalog of northern stars \cite{Lepine2011} to search for super-Earths, as small as twice the radius of 
the Earth, in the Habitable Zone.  GJ1214b, the 
first  transiting super-Earth around a M star, has been identified as part 
of this survey.
\item HATNet \cite{Bakos2002} and HAT-South \cite{Bakos2009}. About 50 M3-M9 dwarfs are monitored per year with HATNet  at a precision per measurement better than
1\%, and 170 per year at a precision better than 2\%, while $\sim$180 M3-M9 dwarfs are monitored per year
with HAT-South   at precision better than 1\%, and 770 M3-M9 dwarfs are monitored per year at
precision better than 2\%. There are
excellent prospects to have a sizeable sample of transiting super-Earth planets around bright (K$\le$9) M-type
stars coming from these and other upcoming ground-based transit surveys.
\item NGTS (D. Pollacco, private communication)
 is based on experience and developments from the WASP instruments and
is designed to allow detection at the 1-2 millimag level around 9th
magnitude stars and 10 millimag at around 15th magnitude. This will allow
the detection of Neptune and SuperEarth sized objects around late type
dwarfs. The main science driver for the NGTS experiment is to understand
the bulk characteristics of these objects. NGTS construction will begin in
spring 2012 at Paranal, Chile.
\item The  Multi-site All-Sky CAmeRA (MASCARA, I. Snellen, private communication), is aimed at finding
the brightest existing transiting planets --Jupiters, Neptunes and super-Earths-- with mag. V= 4-8. 
MASCARA will consist of five identical camera-systems stationed at observatories around the globe. A key element in the design is that from each
station the whole sky will be monitored without the use of any moving parts.
\item
The ground-based transit survey APACHE \cite{Apache} 
is also expected to provide suitable targets for EChO.
This project  is dedicated to the long-term photometric 
monitoring of thousands of nearby M-dwarfs in the Northern hemisphere,  providing the first-ever,  longitudinally distributed network of 
telescopes dedicated to the search for transits of small-size planets.
We foresee the discovery of few tens of super-Earths around M stars
over the next  few years. 
\item HARPS-S (La Silla) and HARPS-N (La Palma) are dedicated radial velocity surveys for exoplanets orbiting bright stars. The preliminary analysis of HARPS data showed that more than 40\% of stars have planets with masses below 50 Earth masses and 30\% of stars with planets below 30 Earth masses \cite{Mayor2009,Lovis2009,Mayor2011}.
\item Further ground-based resources for EChO targets will be radial velocity 
instruments such as CARMENES (Calar Alto, \cite{Quirrenbach}), which was specifically built to search for exo-Earths in the
near-infrared and visible around stars later than M4, with an average 
distance of 15 pc. 
\item Analogously, ESPRESSO \cite{Pepe2010} (Paranal, first light 2016) the next generation of radial velocity instruments 
on VLT, will achieve a radial velocity precision of about 10 cm s$^{-1}$ for stars 
with V $\le$14, which is enough to detect rocky planets in the Habitable 
Zones of late-type stars.
\item While many exoplanets discovered by CoRot \cite{Corot2011} and Kepler \cite{Borucki2011} orbit stars too faint to
allow detailed spectroscopic studies,  broad band photometric observations in the visible and IR with EChO will be feasible for target stars as faint as  V=15 mag, allowing to constrain the equilibrium temperature and the albedo of the planet, or search for exomoons. 
\item  Finally, space mission concepts are currently
considered by NASA, ESA and other national agencies,  and promise to detect additional
transiting planets:   NASA-TESS \cite{TESS}, ESA-PLATO \cite{PLATO} and CHEOPS (CH ExOPlanet Satellite).
CHEOPS is micro-satellite project (35cm size telescope, launch $\sim$2017) in phase A study by a consortium made of Switzerland, Sweden and Austria.
It is designed to detect a transiting Earth-size planet up to periods of 60 days on G, K and early M stars with a V-magnitude brighter than 9th. Mission prime targets include: identified transiting planets, potential transiting planets too shallow to be detected from the ground and stars with a non-transiting hot Jupiter planet. \newline
In the case
these missions are launched, EChO might benefit of an even more enticing
selection of bodies.
\end{itemize}

As mentioned in previous sections, EChO could consider a few non-transiting planets in its target sample  using combined-light observations \cite{Harrington2006}. The aim would be
to study the seasonal changes in 
atmospheric composition/thermal properties in non-transiting eccentric systems
due the significantly variable irradiation conditions: 
for an $e=0.6$ orbit, the stellar flux varies by a factor of 16 along the 
planet's orbit. 
For non-transiting eccentric Neptunes and giants, GAIA will deliver high-quality orbit 
reconstructions and mass determinations, independently of the existence of 
radial velocity 
measurements. If the full orbit and actual companion mass are available, it 
is then possible to predict where and when one will find the planet around 
the star, a key information to study the phasecurve of a non transiting planet.

%Finally, a number of space mission concepts, currently proposed to ESA and
%NASA, promise to detect additional transiting planets: PLATO, TESS, ELEKTRA.
%These missions and EChO are fully independent experiments:
%while EChO might benefit of an even more enticing selection of bodies to choose from, in the case these missions are launched,
%EChO does not require PLATO/TESS/ELEKTRA discoveries to achieve
%the scientific objectives described in the previous sections.

%\subsection{Targets observable, mission schedule{\dots} }
%\remark{Vincent, Ollivier, Micela, Tessenyi}

%\begin{table*}
%\centering\includegraphics[width=0.7\linewidth]{Figures_originals/proposalNov2-img31.png} 
%\centering\includegraphics[width=0.7\linewidth]{Figures_originals/proposalNov2-img32.png} 
%\centering\includegraphics[width=0.7\linewidth]{Figures_originals/proposalNov2-img33.png} 
%\end{table*}

%\clearpage

%% file: payload.tex
\section{Mission concept}
The basic concept for the EChO mission is to build an opto-mechanical system that is as simple as possible whilst fulfilling the requirements set by the scientific observations.  These are dominated by the need to have a system that is as stable as possible to allow repeated observations of transiting planets over periods ranging from the transit time itself, some hours, to the orbital periods of the planets which may be many weeks or months.  The telescope is therefore based around a 1.2-1.5 m primary aperture arranged, in the baseline configuration, in an on-axis Cassegrain configuration.  Behind the telescope we will place a spectrometer that is split into a series of channels to with the goal of covering wavelengths from 0.4 to 16 $\mu$m.  Again different concepts for the spectrometer have been studied and further evolution of the design is likely.  In this section we describe the spacecraft, telescope and instrument conceptual design studies carried out to date and discuss the technical challenges inherent in the design and realisation of the EChO mission.

\subsection{Spacecraft Overview} 
The EChO system design is driven by the stability and sensitivity requirements of differential spectroscopy in the optical, near and thermal infrared. The spacecraft driving design requirements are: (1) Provision of cooling for the telescope 
and the spectrograph to $\sim$50 K  and $\sim$45 K (30 K for the detectors), 
respectively, (2) high photometric stability requirement which translates to 
line of sight stability and payload thermoelastic environment stability 
requirements; and (3) overall compatibility with cost and technology readiness criteria of an ESA M-class mission (see http://sci.esa.int/science-e/www/object/index.cfm?fobjectid=47570), i.e. launch on a Soyuz-Fregat rocket and cost cap to ESA of 470MEuros.
\begin{figure}[h] 
\centering
\includegraphics[width=0.4\textwidth]{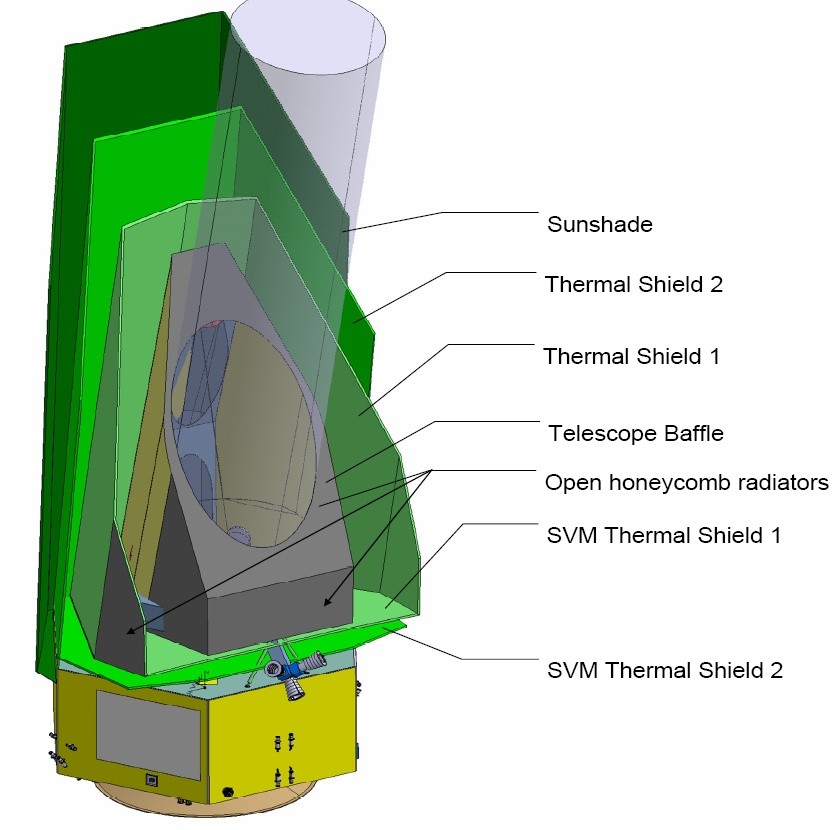}
\includegraphics[width=0.3\textwidth]{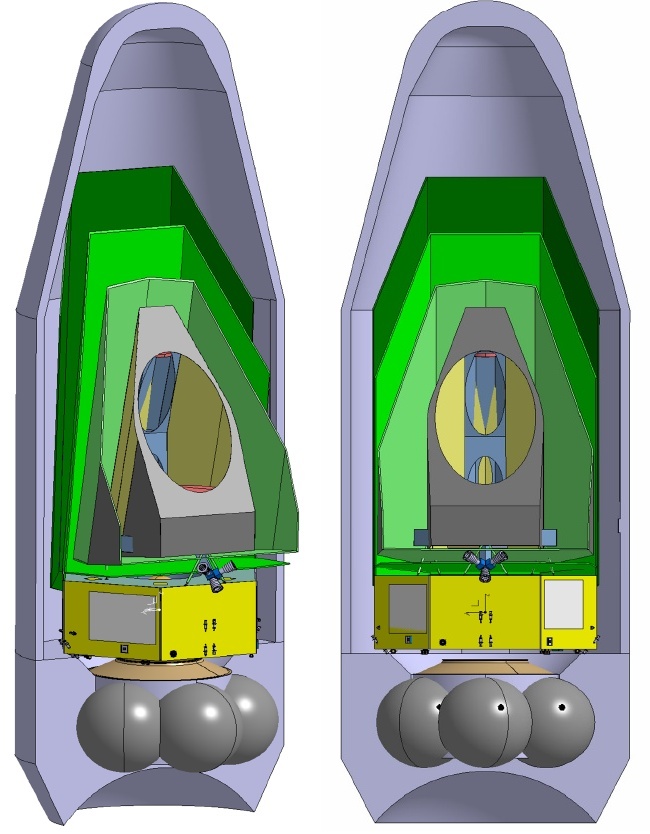}
\caption{Top: Baseline design for EChO thermo-mechanical design concept. 
Bottom: EChO spacecraft baseline configuration under the Soyuz-Fregat fairing.}
\label{figure_global_EChO}
\end{figure}  

The EChO spacecraft will consist of three major modules (fig. 
\ref{figure_global_EChO}): (1) A payload module (PLM) including the instrument (telescope hardware, focal plane assemblies) containing all functionalities required for the scientific measurements. (2) A service module (SVM) containing all functionalities and equipment required to control the payload and support the scientific observations. The SVM accommodates all the spacecraft subsystems such as propulsion, communication, power, and the attitude and orbit control system (AOCS). (3) A multi-layer sunshield for protecting the PLM from solar irradiance thus providing the required thermal environment for the telescope and instrument.  

\subsection{Model payload}
\subsubsection{Overview of the payload}
\begin{figure}[h] 
\centering
\includegraphics[width=0.5\textwidth]{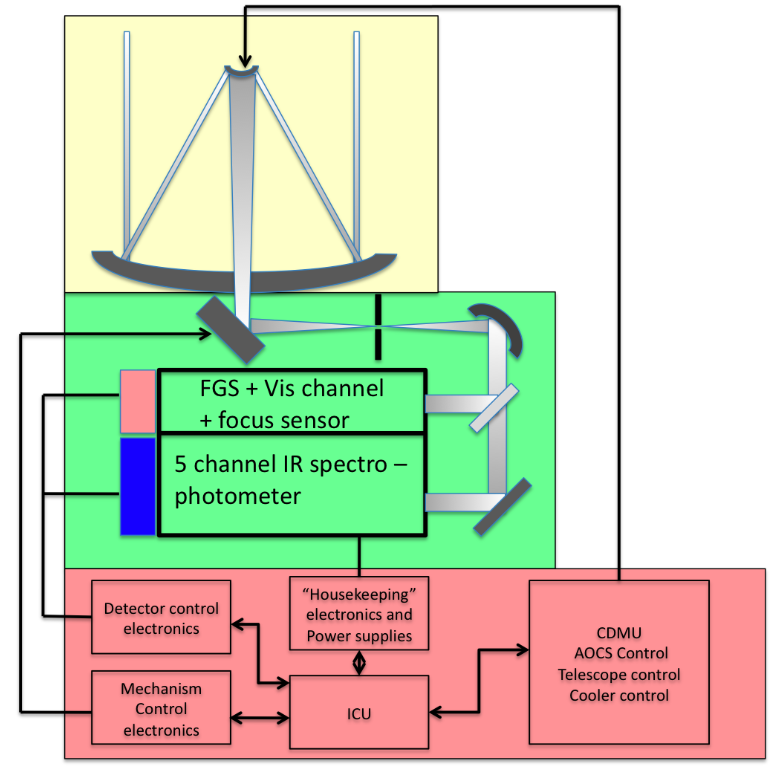}
\caption{Scheme of the payload. Maximum temperature colour code is : yellow $\sim$ 50~K, 
green $\sim$ 45~K, blue $\sim$ 30~K, red $\sim$ 300~K.}
\label{figure_global_instrument}
\end{figure}

In the baseline concept, the payload (Fig.~\ref{figure_global_instrument}) consists of a single 
instrument, a multi-channel spectrograph, at the focus of a 1.2-1.5~m effective aperture telescope.
In this configuration, the spectrograph covers the spectral range from 0.4 to 16 $\mu$m with one visible to near-infrared (0.4 -- 1.0 $\mu$m, hereafter the visible channel) and five near to mid-infrared channels.  The spectral resolution will vary from a few tens in the thermal infrared to a few hundreds in the near infrared.  The visible channel is used for stellar activity monitoring and will also contain a fine guiding sensor (FGS) and focus sensing (FS). One possible implementation for the FGS will be to employ a tip-tilt mirror to stabilise the stellar image on the common input field of view for the spectrographs.  Another possibility is to use the spacecraft attitude control system directly; both options are subject to continuing study. 

The telescope temperature should not exceed 50~K, with
the instrument structure and mirrors at 45~K and the detectors at 30~K. The telescope and instrument structure can be cooled using passive techniques, but, due to the low temperature and high stability requirements, mechanical coolers must be used for the infrared detectors.  

The instrument electronics will consist of a digital processing unit (denoted ICU in figure 11), a set of signal conditioning electronics for the detectors, a second set for the thermal sensors etc and the drive electronics for the proposed tip~tilt mirror in the fine guidance system.  It is possible that only a single on-board computer and data storage unit will be employed for the spacecraft which will also control the payload.  In this case the instrument conditioning electronics will act as a remote terminal unit with limited processing capabilities.  Consideration of this configuration is made possible due to the relatively small data volume from the payload, we estimate no more than $\sim$3.5 GBits/day including housekeeping and margins.

\subsubsection{Instrument baseline conceptual design and key characteristics}
\begin{figure}[h]
\includegraphics[width=0.5\textwidth]{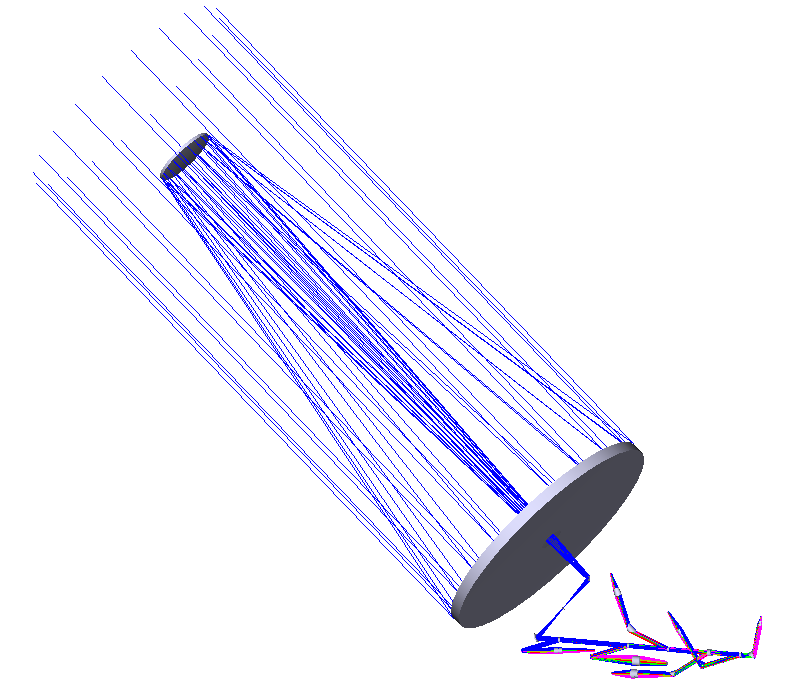}
\includegraphics[width=0.5\textwidth]{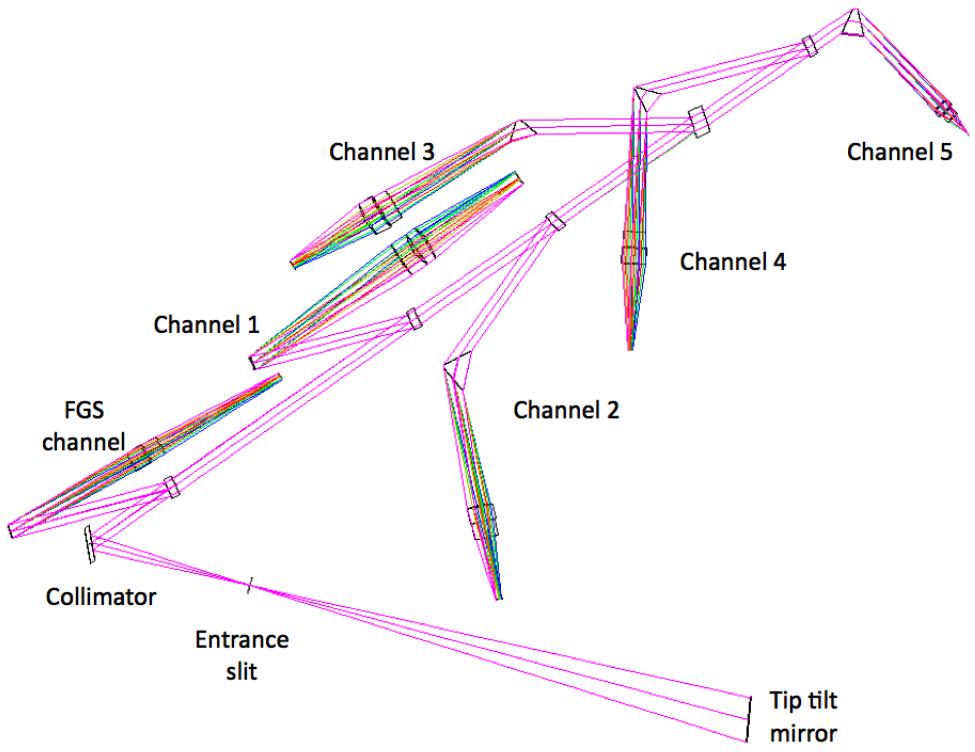}
\caption{Left: \textbf{Baseline optical concept} with on-axis telescope and full diffractive/dispersive spectrograph.
Right: \textbf{Baseline optical design} of a dispersive implementation of the instrument. 
The FGS channel is also a FS channel and a 0.4-1 micron spectrograph, with a
spectral resolution of few tens to few hundreds, to monitor the stellar activity. }
\label{figure_on_axis_concept}
\label{fig_prelim_opt_design}
\end{figure}

A classical on-axis Cassegrain telescope
with a $\sim$1.5~m class primary mirror diameter  would meet
the main requirements of EChO such as cooling to 50~K by passive techniques
and reaching the PSF quality requirements within a central small field
of view (several arcmin) around the target star. The design and development
of such a telescope shown on Fig.~\ref{figure_on_axis_concept} does not present any particular difficulty.  Due to its stability requirements at 50~K operating temperature silicon carbide (SiC) ceramic \cite{Pilbratt2010} would be the material of choice for the telescope, given its high thermal conductivity and high mechanical stability at cryogenic temperatures compared to Zerodur/CFRP or Zerodur/ceramic technology. Another advantage of SiC for EChO is the absence of moisture release associated with CFRP.

A dispersive/diffractive spectrograph has been investigated as baseline
concept. The spectrograph is fed directly by the telescope image through an entrance aperture
whose size is slightly bigger than the PSF size at 16 $\mu$m.  As a consequence,
the entire flux of the star is used in the different channels and no slits are used.  A preliminary design of this concept has been done and is presented on Fig.~\ref{fig_prelim_opt_design}). The
characteristics of each spectral channel is presented in table \ref{table_prelim_opt_design} assuming
the requirements on spectral channels (spectral range, spectral resolution) and the characteristics of available detectors (size, pixel size and number).

One of the key-points for the instrument design, which drives its development, is the choice of the detectors. This choice is strongly constrained by the availability in Europe of very few detectors in the 0.8 -- 16 $\mu$m
spectral range. The best European detectors in this spectral
range are based on InSb (NIR) and HgCdTe (MCT) technologies. For homogeneity
of the focal plane and readout electronics, we have based the 
prototype design exclusively on MCT photovoltaic
detectors.  These are available in the form of 350$\times$250 or 500$\times$500 pixel matrices from several suppliers.  The current generation of n on p devices should meet the EChO requirements up to about 11 $\mu$m if they are operated at temperatures of about 30 K \cite{Terrier2010,Itsuno2011}. The detector for the VIS/NIR channel (0.4 -- 1 $\mu$m), FGS and FS is a
classical CCD, commercially available in Europe\footnote{e.g. http://www.e2v.com/products-and-services/high-performance-imaging-solutions/space---scientific-imaging/}. 

Technological improvements are necessary for the 11 -- 16  $\mu$m
spectral range to reduce the level of dark current in currently available MCT detector arrays. An alternative could be the use of Si:As (IBC) technology, leading to
virtually ``noiseless" detectors.  This technology was used on Spitzer and will be flown on JWST 
\cite{VanCleve1995}. However, present products work at 7~K, require long readout time to keep the noise level low and exhibit low full well capacity.  The latter leads to the need for frequent readout to maintain performance and, therefore, to a reduction in the overall duty cycle of the instrument.  The need for low temperature operation, in particular, will have a large impact on the overall mission design, most especially the need for closed cycle mechanical coolers operating at up to 7~K.  Solutions do exist for these in Europe \cite{Planck} 
and  this option will be studied extensively as part of the ongoing mission design exercise.

\begin{table}[h]
\rowcolors{1}{blue!10}{}
\tiny
\begin{tabular}{|p{1.1cm}|p{0.8cm}|p{0.8cm}|p{0.8cm}|p{0.8cm}|p{0.8cm}|p{0.8cm}|}
\hline 
~ &{ FGS + FS+} Vis/NIR channel & Channel 1 & Channel 2 & Channel 3 & Channel 4 & Channel 5\\\hline
 Bandpass ($\mu$m) & 0.4-1 & 0.8-2.7 & 2.3-5.2 & 4.8-8.5 & 8.3-11  & 11-16\\ \hline \hline
% \multicolumn{7}{c|}{ }\\
  \multicolumn{7}{c|}{   \small{TELESCOPE}}\\ \hline
 Diameter &\multicolumn{6}{c|}{  1.2 to 1.5 m. 1.4 m telescope used here}\\\hline
 F\# & \multicolumn{6}{c|}{10}\\\hline
 Transmission &\multicolumn{6}{c|}{98\%}\\\hline \hline
%  ~ & \multicolumn{6}{c|}{ }\\
 \multicolumn{7}{c|}{  \small{COLLIMATOR: off-axis parabola}}\\\hline
 Focal length & \multicolumn{6}{c|}{  200 mm}\\\hline
 Transmission & \multicolumn{6}{c|}{  99\%}\\\hline \hline
% & ~ & \multicolumn{6}{c|}{ }\\
 \multicolumn{7}{c|}{\small{OBJECTIVES}}\\\hline 
 Type & Doublet  & Doublet  & Doublet  & Doublet  & Doublet  & Doublet\\\hline
 Material & { PHH 71}

                   LAH 54 &{ F\_Silica}

                                     CaF2 & ZnSe 
                                                 
                                                 Germanium &{ Silicon }

                                                                        AMTIR 1 & { Silicon}

                                                                                            AMTIR 1 & { CdTe}

                                                                                                                  CdSe\\\hline
 Focal length & 200 mm & 150 mm & 100 mm & 100 mm  & 100 mm  & 50 mm\\\hline
 Image F/\# & 10 & 7.5 & 5 & 5 & 5 & 2.5\\\hline
 Transmission & 95 \% & 95 \% & 95 \% & 95 \% & 95 \% & 95 \%\\\hline  \hline
%  ~ & \multicolumn{6}{c|}{ }\\
 \multicolumn{7}{c|}{\small{DISPERSION SYSTEM}}\\\hline  
 Type  & Grating & Grating & Prism & Prism & Prism & Prism\\\hline
 Grating density &111/ mm & 64 /mm & N/A & N/A & N/A & N/A\\\hline
 Material & N/A & N/A & CaF2 & CaF2 & Cleartran & CdTe\\\hline
 Prism angle & N/A & N/A & 62 Â°  & 47 Â°  & 59 Â° & 59 Â°\\\hline
 Spectral resolution & 600 & 600 & 600 or better (to be studied) & 600 & 600 & 20\\\hline
 Transmission & 60 \%  & 40 \% & 90 \% & 90 \% & 90 \% & 90 \%\\\hline \hline
%  ~ & \multicolumn{6}{c|}{ }\\
\multicolumn{7}{c|}{\small{DETECTOR}}\\\hline  
 Type & CCD &{ HgCdTe}

 SWIR &{ HgCdTe}

 MWIR &{ HgCdTe}

 LWIR & HgCdTe
 
 LWIR &{ HgCdTe}

 VLWIR\\\hline
 Pixel size & ~ & 30 $\mu$m & 15 $\mu$m & 30 $\mu$m & 30 $\mu$m & 30 $\mu$m\\\hline
 Needed pixels & 600 & 650 & 460 & 330 & 182 & 40\\\hline
 Working temperature & ~ & {\textless} 110 K & {\textless} 80 K & {\textless}40 K & {\textless} 40 K & 30 K\\\hline
 Quantum efficiency & ~ & 0.5 & 0.7 & 0.7 & 0.7 & 0.7\\\hline
 Dark current (e-/s/px) & ~ & {\textless} 10$^{(1)}$ & {\textless} 10$^{(1)}$ & 500 & 500 & 10000(2)\\\hline
{ Readout noise }

 (e-/px/ro) & ~ & 150 & 400 & 1000 & 1000 & 1000\\\hline
\end{tabular}
\caption{Main characteristics of instrument and detectors.
Objective and prism materials are detailed for the dispersive solution.}
\label{table_prelim_opt_design} 
$^{(1)}$ This performance measured at temperatures lower than standard
operating temperature\\
$^{(2)}$ Expected performance after technology development programme
\end{table}

\subsubsection{Pointing and alignment requirements}
Another key design driver for the instrument is maintaining the stability of the line of sight, i.e. the stability of the 
stellar image at the entrance aperture of the spectrograph.  This is required in order to maintain the photometric stability 
over the spectra and avoid inter and intra pixel inhomogeneities. The spectrograph is
designed so that the size of the
PSF seen through the spectrograph is about one pixel on each spectral channel. The required stability of the line of sight will be a fraction of the PSF at the shortest wavelength, this translates into about 20 mas at 1 $\mu$m.  EChO is not an imaging
mission and therefore we do not need to stabilize the rotation of the field but only the line of sight for an on-axis target.
One option for stabilisation of the line of sight is to use a combination of the of the spacecraft Attitude and Orbit Control System (AOCS), which will have absolute pointing of several arcsec and stability of several hundreds of mas, and a fast steering mirror within the instrument (see \S\ref{sec_aocs}).  The latter will stabilise the position of the stellar image at the entrance aperture with an accuracy of several tens of mas with a response frequency of order of a few Hz.

The stringent spectro-photometric stability requirements for EChO require a 
stable thermal and thermoelastic optical system. 
The alignment tolerances are 
estimated to be of the order of 2 $\mu$m for a telescope compatible with visible 
light operation down to wavelengths of a few hundred nm.  Given the inability to predict the behaviour of the telescope when operational in flight --i.e. the detailed effects of gravity release, thermal contraction etc.-- we expect to have to re-focus the telescope once it has achieved a stable temperature. In the baseline, on-axis, configuration, the in flight focus of the telescope is determined by the use of a focusing device adjusting
the position of the telescope secondary mirror. This device is driven
by a focus sensor within the VIS/FGS unit that analyses the size of the
image at the entrance aperture of the spectrograph.

\subsubsection{Calibration and other specific requirements}
\label{calibration}
The basic measurement technique involves extracting spectrally separated signals in phase with the known period of our target planets. The difference between the stellar signal and the planet is very small (in the extreme $\sim$1 part in 10$^5$) and in some cases, many transits are required in order to build sufficient signal to noise to extract a high fidelity planetary only signal.  Clearly accurate calibration and signal stability are going to be of the utmost importance both during and before and after each transit observation and in attempting to combine observations.
\begin{figure}[h]
\includegraphics[width=0.45\textwidth]{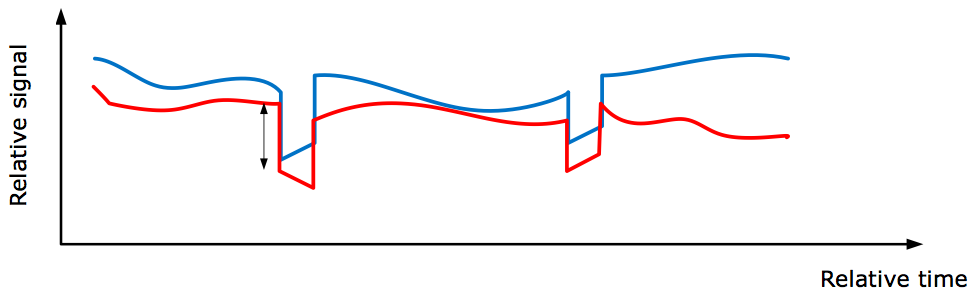}
\includegraphics[width=0.45\textwidth]{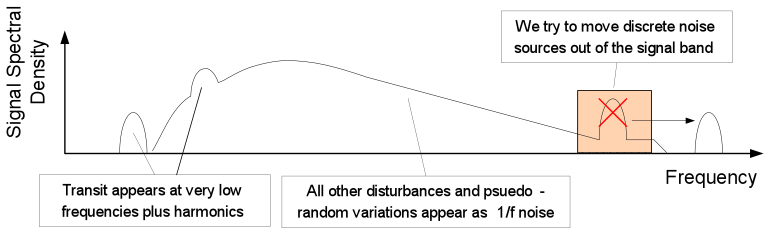}
\caption{Top: Measurement situation in the time domain. With (exaggerated) differences 
between two visits to a transiting planet.  The period of the transit plus stellar flux 
measurement determines the lower bound of the instantaneous
signal band - the upper bound is defined by the electronics low pass filter.  
The period between transits defines the frequency at which the primary signal 
is extracted and the time between visits defines the lower band of interest in the frequency space.
Fourier decomposition and accurate calibration of all possible sources of disturbance 
are required to avoid systematic uncertainties at the extraction frequency. \newline
Bottom: Measurement situation rendered in frequency space.  The shaded box illustrates 
the instantaneous frequency band of the measurement and all high frequency noise 
sources should be kept away from this band.  The primary signal extraction frequency 
is at very low frequencies and the EChO calibration and observation stability must be 
such as to avoid allowing unknown systematic uncertainties into this frequency space}
\label{fig_calib_1}
%\label{fig_calib_2}
\end{figure}
The general situation is illustrated in Fig.~\ref{fig_calib_1}
indicating how the variations 
due to the (roughly) square wave transits are transformed into a pseudo monochromatic 
spectral feature in frequency space plus higher harmonics.  The fluctuations due to 
variations in stellar flux, temperature drifts, detector drifts, miss-pointing and 
jitter etc are super imposed in frequency space as some form of 1/f ``pink" noise spectrum with (hopefully) a low frequency cut off somewhere above the primary frequency band for the transits.  Some sources of noise, such microphonics from reaction wheels or mechanical coolers, may lead to higher frequency disturbances which, if not removed, will add as white noise.  One way of combating these is to ensure that they are at frequencies above the low pass filter cut off in the detector signal chain.  

We anticipate that most sources of uncertainty will either enter as white noise (accounted for in the sensitivity analysis) or at frequencies sufficiently removed from the signals of interest that they can be ignored or removed by correlation.  
Some disturbances will, however, be in phase with the measurements and we will need to calibrate certain aspects of the systems with extreme care. 
We need to consider the following in some detail:
\begin{itemize}
\item The ability to point the target at exactly the same location onto the instrument and any variation in instrument response across its field of view.  Miss-pointing when re-acquiring the target is in direct phase with the measurement frequency and will be difficult to deal with as it requires a stable astronomical source which cannot be guaranteed for our target stars 
(see below). This aspect can only be assessed and calibrated with multiple pointing at stable sources associated with highly accurate calibration measurements of the spatial response of the system.
\item High frequency variation in signal response due to detector instabilities or satellite disturbances.  These will not necessarily be removed by the phase detection and will add to white noise in the measurement.  Unless these are correlated with other measurements 
on the spacecraft (temperature, reaction wheel frequency etc) they are difficult to remove.  
This means that careful attention should be made on where the frequencies of possible disturbing elements are placed and all possible correlating sensors (thermistors, position indicators, voltages etc) must be sampled with sufficient fidelity that they can be used to perform de-correlation of the signals, if necessary, in ground processing.
\item Any stellar variation on timescales similar to the period of the transiting object that could mimic a transit signal.  This may be unlikely but still needs to be monitored and considered in detail.
\item Variation in the instrument response over a long timescale must be monitored as this will also lead to systematic variations in the signal at the frequency of interest.
\end{itemize}
In general attempting to provide absolute calibration of the system to the level of 1 part in 10$^5$ is extremely difficult and ultimately limited by our lack of detailed knowledge of the characteristics of stellar variability at this level.  Rather our strategy will be to 
eliminate as much of the systematic sources of uncertainty as possible following the discussion in this section and rely on the target stars themselves, in conjunction with an on board calibration source, to provide repeated measurements of the instrument behaviour over many timescales thus providing longitudinal calibration trends to remove all possible sources of systematic uncertainty.  A highly stable and repeatable calibration source is an essential part of this scheme and our outline design envisages placing a source at the location of the telescope secondary where (in an on-axis design at least) the anti-narcissus beam dump (hole) will be placed at the centre of the mirror. Placing the source here, has the major advantage that the calibration source is viewed through as much of the optical chain as possible, and therefore monitors the total degradation in performance over the course of the mission. Our initial goal for the calibration source is that it is absolutely stable to 1 part in 10$^3$ over at least one year of operation, with a rather higher short term stability to monitor the system during the period of each transit.  The absolute output of the source will be verified periodically against the most stable sources available.  The entire calibration scheme and the requirements on the various components of the system require detailed study during the assessment phase and will be the subject of a future report.

\subsubsection{Current heritage and Technology Readiness Level (TRL) for the payload}
Current heritage used for the instrument of EChO and Technology Readiness Level of subsystems are summarised in table \ref{table_TRL_instrument}. 
\begin{table}[hbt]
\rowcolors{1}{blue!10}{}
\begin{tabular}{|p{2.5cm}|p{2cm}|p{2.5cm}|}
\hline
~  & TRL & Current heritage / \newline development\\\hline
{ European HgCdTe detectors} SWIR/MWIR & 8 & Planetary missions: MEX, Phobos Grunt\\\hline
{ European HgCdTe detectors} LWIR/VLWIR with low dark current & 3 at present, should be TRL 5 within the next 2 years& ~ \\\hline
{ American  HgCdTe/SiAs} detectors LWIR/VLWIR with low dark current & 8 & Spitzer, JWST  \\\hline
 Optimised readout circuits & 3 (24 to 36 months required to reach TRL 6)&  \\\hline
 Dispersive spectrograph / optics & 8 & Heritage IR spectrographs \\\hline
 FTS spectrograph knowledge / technology& 5 (prototypes) to 9 (instruments 
 already launched) & SPIRE(Herschel) / IASI (Metop) / MIPAS (Envisat) \\\hline
Detectors electronics chains  & 5 & \\\hline
DPU and on board processing electronics & 9 & \\\hline
Fine guiding system & 5 & \\\hline
\end{tabular}
\caption{Technology Readiness Level (TRL) and current heritage for payload subsystems}
\label{table_TRL_instrument}
\end{table}

\subsubsection{Attitude and orbit control}
\label{sec_aocs}
The EChO spacecraft needs to be 3-axis stabilized in order to be able to provide a high instrument line-of-sight pointing accuracy and stability of few tens of milli-arcsec (mas).  As design criterion, the size of the PSF diffraction limited at the short-wavelength end of the infrared spectrograph of 0.8 $\mu$m is 110 mas for a 1.5~m telescope and 35 mas is about a quarter of its size. Using a dispersive spectrograph with entrance slit, successfully demonstrated on Hubble and Spitzer, as worst-case scenario requires an RPE of 30 mas over 500 s.

To avoid stringent pointing requirements on the AOCS alone we propose separating the overall satellite AOCS and instrument line of sight control by using a fast fine-steering/tip-tilt mirror within the instrument.
This is possible since the main science objectives are related to the central stellar source and corrections for the central source can be performed by a movable pick-up mirror located before the beam splitter. In particular, the field rotation in the focal plane due 
to AOCS fluctuations has no consequence on the performance of the instrument because the only requirement is the stability of the on-axis PSF at the spectrograph entrance aperture.

For the overall AOCS the requirement on the relative pointing error (RPE) is 500 milli-arcsec over 500 s with an absolute pointing error (APE) of a few arcsec.  This can be achieved with classical reaction wheels controlled by star trackers which have to be mounted under optimised thermoelastic constraints onto the payload module. The RPE is dominated by wheel noise in the frequency domain up to 200 Hz. Since our target stars are bright in the optical, the bandwidth of the control signal for the fine control of the fine-steering mirror can be sufficiently high.  The feedback for the position control is provided by a Fine Guidance Sensor (FGS) included in the instrument visible channel which will monitor the position of the star image at the spectrograph aperture entrance.  The FGS information may also be used by the global AOCS loop.

\subsubsection{Thermal aspects}
\label{sec_thermal}
\paragraph{Baseline design}
Cooling the telescope and the instrument to cryogenic temperatures between 45 to 50 K requires an effective sunshade/sunshield system shading the payload from the sun and insulating it from the warm SVM. Furthermore the bipods supporting the telescope optical bench (TOB) have to have low thermal
conductivity and long conductive paths.  Low conductivity materials (manganin, phosphor
bronze, stainless steel) are also necessary for the harness between the SVM and the telescope and instrument. The total parasitic thermal load has to be reduced to a level that can be balanced by the heat rejected from the payload radiating surfaces (T $<$ 50 K) towards deep space.

For the sunshade/sunshield system different concepts will be traded in the concept study taking into account the following major constraints:
\begin{itemize}
\item The sunshade/sunshield system has to be accommodated under the Soyuz-Fregat fairing.
\item The sun -- spacecraft -- line-of-sight angle shall be large enough. Goal is to achieve a solid angle of 1
to 2{$\pi$} steradian at any time.
\item The structural design of the sunshade/sunshield system shall be compatible with mass and mechanical requirements.
\item The thermal design shall allow the telescope to be passively cooled to T $<$ 50 K.
\item The use of structural material and MLI shall be compatible with outgassing requirements.
\end{itemize}
Different design concepts have been considered. Herschel uses a single
sunshade/sunshield system and a separate single SVM thermal shield to passively cool the liquid He cryostat
vessel to 70--80~K. This is insufficient for EChO and for this reason a multiple system consisting of a sunshade and two separate thermal shields is suggested taking advantage of the V-groove effect as successfully applied to the Planck-spacecraft \cite{Planck}. 

\subsection{Orbit choice and baseline mission scenario }
\subsubsection{Orbit}
Given the need to cool the payload and maintain a stable thermal 
environment the choice of orbit is limited to the Earth trailing type such as 
used by Spitzer or the second Lagrangian point (L2) Lissajous (PLANCK and Herschel) or the L2 halo orbits.  
The Sun synchronous low Earth orbit, which may have the 
benefit of allowing a larger mass and therefore larger mirror, has many 
drawbacks in terms of observing modes. Given that the Earth trailing orbit has 
a disadvantage in terms of a decreasing data rate capability as the satellite 
drifts away from the Earth, we are assuming that the satellite will be placed 
into a halo orbit at L2 with a large radius (800 000 km). 
The large radius halo orbit is preferred as it reduces the sun-satellite distance variation 
speed along the terrestrial revolution and avoids eclipse over the complete mission
with no significant $\Delta$V requirements, greatly simplifying the thermal design and 
spacecraft operations. In addition, it  allows a greater mass for the satellite. 
We estimate that the overall mass of the system presented here will be less than 2.1 T allowed
for a Soyuz Fregat launcher. 

\subsubsection{Baseline operational scenario}
\label{mission_operational_scenario} 
The typical mission duration is 5 years. The nature of the scientific mission 
that EChO will carry out is to visit a restricted set of stellar targets ($\sim$100) at prescribed times to match the orbital phase of the exoplanet.  
Typical visits per target will be of the order of 1--10 hours 
interspersed by periods of a few to up to 20 days.  Given that target stars 
may be in any part of the sky and will have a range of orbital periods it is 
essential that the satellite has a large ``field of regard", i.e. as little 
restriction as possible on the direction the satellite can be pointed due to 
Solar and terrestrial viewing constraints. This requirement will, in turn, 
drive the basic direction of spacecraft pointing and the arrangement and size 
of the thermal shields, antennae and Solar arrays. A typical angle of $\pm$ 30 degrees with the Sun-Earth normal appears to be a good trade-off (cf. 
observation strategy section). It allows a full visibility of the sky (all the targets) over one year, and an instantaneous visibility of more than 40\% of the targets.
The observation plan can thus be adapted according to the scientific interest of the targets.

Given the predictable nature of the observing plan we envisage that the 
satellite will be able to carry out semi-autonomous operations for reasonably 
long periods of time with minimal ground contact required (typically 2 per week).  
During the observing periods we anticipate that the 
majority of the science data will be stored on board in a suitably compressed format.  The high rate transfer of the science data to the ground will be undertaken during dedicated ``Data Transfer and Commanding Periods" (DTCP) in a similar manner to the Herschel and PLANCK operations but with a rather lower duty cycle.

\subsection{Alternative design studies of key items}
Alternative options to the baseline concept have been considered.
The payload and mission requirements have been studied 
together with the industry (Astrium GmbH, Germany, Astrium UK) 
and national agencies (CNES) to confirm technical feasibility of key items within the M-Class 
cost cap. A detailed design trade-off will to be performed during the assessment study.
\paragraph*{Telescope -- } ~
The need for EChO's spectrophotometric stability of 10$^{-4}$ over several hours
requires a thermally stable telescope, i.e. a well baffled and shielded system with high
mechanical stiffness and stability. 
These design drivers led to the possibility of a 1.5\,m off-axis design
(1.45\,m effective diameter) which has been investigated in order to optimize
the thermal shielding. 
The off-axis system with M2 location close to the thermal shields (see Fig.~\ref{figure_global_EChO}) allows
implementation of an oversized shield which provides high thermal stability since mounting spiders
of an on-axis secondary mirror can be avoided, minimising drifts in thermal emission due to temperature changes.
This design also has no energy loss due to clipping from the central
obscuration and also provides a cleaner pupil. 
This concept easily accommodates
beam-splitting into a visual channel including the FGS (followed by a relay
optics towards the star sensor chip) and into the scientific IR channels 
towards the spectrometers including a focus sensor (e.g. Shack-Hartmann or specific device).

\begin{figure}[h]
\centering
\includegraphics[width=0.4\textwidth]{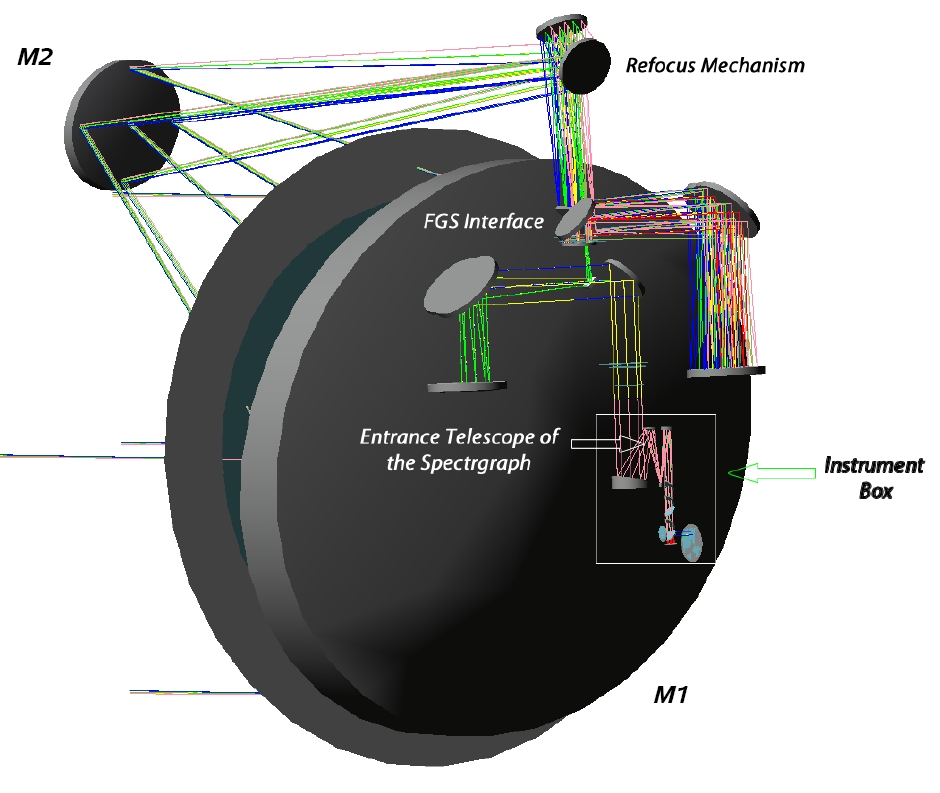}
\includegraphics[width=0.5\textwidth]{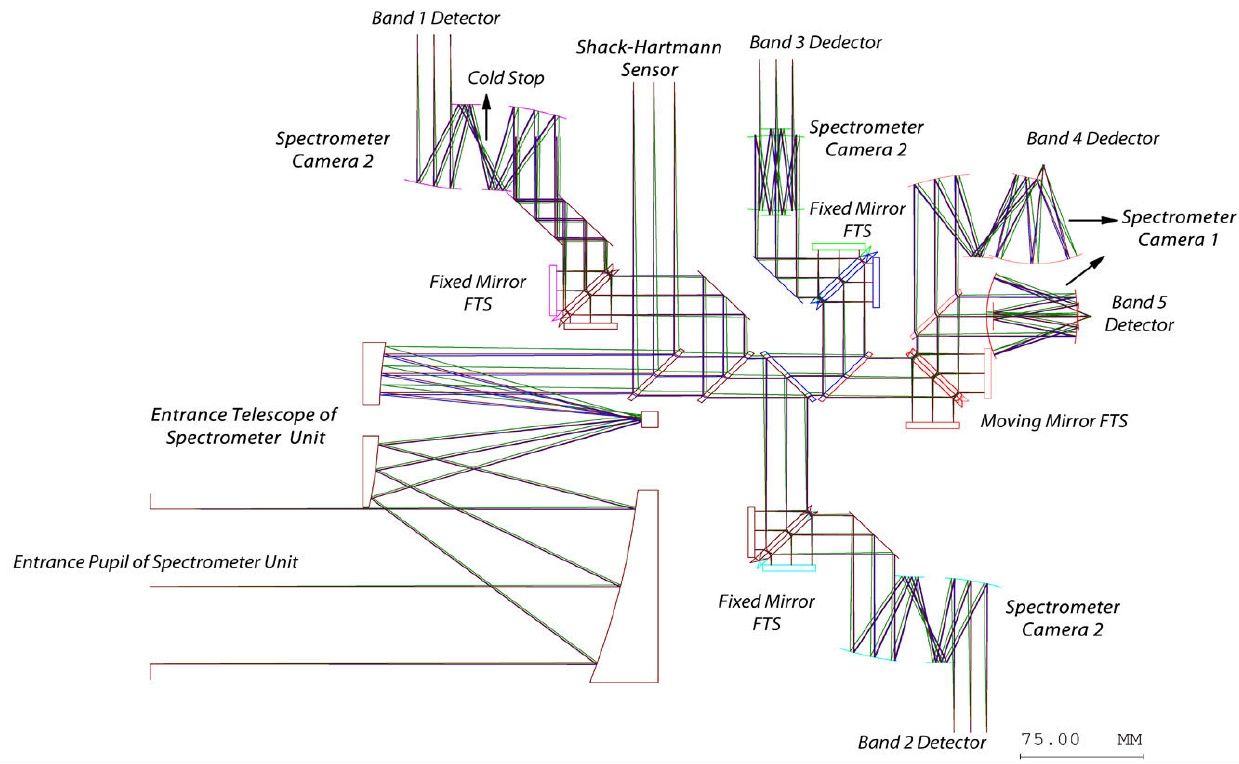}
\caption{Top: alternative design proposed for a 1.5\,m off-axis telescope (Astrium Germany): 
3D view of the payload optical path.
Refocus and FGS units are also indicated. \newline
Bottom: Alternative   spectrograph proposal
using Fourier transform spectrometers (FTS)
for all channels. A series of dichroics separates the beam into the
different spectrometer channels. Band 1-3 are realised with fixed-mirror FTS 
using anamorphic cameras in reflective optics for imaging onto the detecors.
The long-wavelength channels 4 and 5 are implemented as scanning FTS. 
}
\label{fig_prelim_opt_design_fts}
\end{figure}

\paragraph*{Instrument -- } ~
In addition to the dispersive/diffractive design an implementation of the spectrometer channels based on Fourier transform schemes has been studied. Fourier transform spectrometers (FTS) come in a variety of designs including the use of gratings in the interferometer arms (so called Spatial Heterodyne Spectrometer - SHS) or using a beam shearing system. In the FTS implementation studied for the EChO
instrument (see Fig.\ref{fig_prelim_opt_design_fts}) all channels are behind
the collimator which reduces the beam size of the scientific path from
100 mm down to 25 mm. A series of dichroics separates the beam into the
different spectrometer channels.  The spectral bands covering from 1 to 8 $\mu$m (1, 2 and 3) are implemented
with fixed-mirror FTS with the interferogram generated by tilting one
mirror according to the required resolution. The image is focused by an
anamorphic camera in reflective optics onto an approximately 1024$\times$10 detector array. The pupil image is reduced to the size of 20 mm in the afocal direction on which the spectrum
is spread. The focal direction reduces the spot size form of the star
image onto a few pixels. Conservative assumptions on the dark current of currently available European detectors does not permit a fixed mirror FTS with a detector array in CMOS
technology. A scanning FTS with a single or few pixel detector is therefore proposed for the two long wavelength channels. 

\paragraph{Alternative thermal designs}
\begin{figure}[h]
\centering\includegraphics[width=0.3\textwidth]{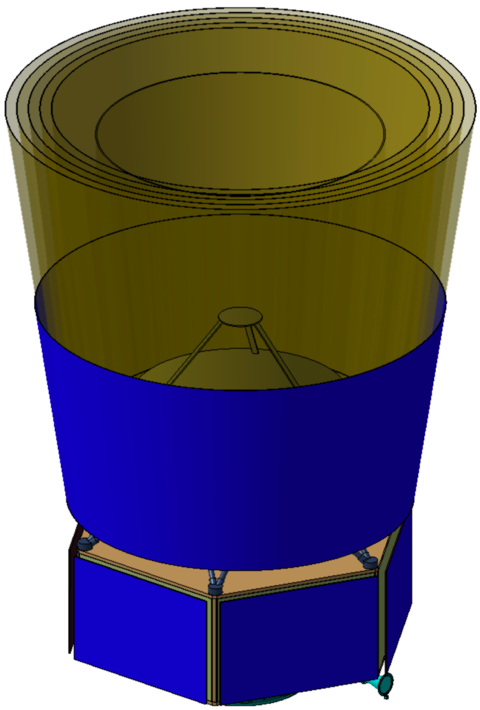}
\caption{Alternative thermal design with curved shields proposed by Astrium UK to allow an increase of the observable sky fraction.}
\label{fig_curved_shield}
\end{figure}
Alternative thermal designs were considered (Astrium UK) to allow an increase of the roll angle with respect to the position of the satellite on its orbit in the 
antisolar direction and as a consequence, an increase of the sky fraction, which is instantaneously visible.
Such concepts, circular or semi-circular sunshields, require the use of curved panels as illustrated on Fig.\ref{fig_curved_shield}.

A preliminary study of these concepts showed that they are slightly more complex, with a small  mass increase, compared to the baseline solution. However, they still maintain comfortable margins with respect to Soyuz-Fregat launch capabilities.
A complete payload systems design trade-off, including scientific performance, will be undertaken during the assessment phase.
%%%%%%%%%%%%%%%%%%%%%%%%%%%%%%%%%%

%% file: conclusions.tex
\section{Conclusions} 
We have presented in this paper the Exoplanet Characterisation Observatory --or EChO-- which will 
provide an unprecedented view of the atmospheres of planets around nearby 
stars  searching for molecular features. 
Those planets will span a range of masses (from gas giants to super-Earths), stellar companions (F, G, K and M) and temperatures (from hot to habitable). 

 EChO will inherit the technique and exquisite photometric precision of CoRoT and Kepler, aiming at the 10$^{-4}-10^{-5}$ level of precision
in the observation of the target-star at multiple wavelengths. 
 The scientific wisdom of having a broad wavelength coverage from the optical to the IR (0.4 to 16\,$\mu$m) comes from nearly 50 years of remote sensing observations of planets in our Solar System combined with  the more recent experience of (very) remote sensing of exoplanetary atmospheres.
EChO will observe the atmosphere of planets already discovered by other surveys and facilities. If launched today, EChO would select $\sim$80 targets for atmospheric characterisation
out of the almost 200 confirmed transiting exoplanets. Most of these targets were discovered by dedicated ground-based transit/radial velocity search programmes, which are
 presently delivering a flood of new discoveries. A  new generation of transit/radial velocity surveys (MEarth, APACHE, HATSouth, NGTS, CARMENES, ESPRESSO etc.) will provide
 new access to the population of  Earth-mass planets orbiting bright late type-stars, e.g. GJ 1214b. 

Finally, in the quest for habitable worlds outside our Solar System, EChO will be able to observe, among other targets,  super-Earths  in the temperate zone of M dwarfs: these are clearly not the Earth's and Sun's \emph{twins}, but rather their \emph{cousins}. Will  they present equal opportunities for habitability?

%% file: article_v1.bbl
\begin{thebibliography}{10}
\providecommand{\url}[1]{{#1}}
\providecommand{\urlprefix}{URL }
\expandafter\ifx\csname urlstyle\endcsname\relax
  \providecommand{\doi}[1]{DOI \discretionary{}{}{}#1}\else
  \providecommand{\doi}{DOI \discretionary{}{}{}\begingroup
  \urlstyle{rm}\Url}\fi

\bibitem{Cho2003}
J.~{Cho}, K.~{Menou}, B.M.S. {Hansen}, S.~{Seager}, ApJ \textbf{587}, L117
  (2003).
\newblock \doi{10.1086/375016}

\bibitem{Thrastarson2010}
H.T. {Thrastarson}, J.~{Cho}, ApJ \textbf{716}, 144 (2010).
\newblock \doi{10.1088/0004-637X/716/1/144}

\bibitem{Harrington2006}
J.~{Harrington}, B.M. {Hansen}, S.H. {Luszcz}, S.~{Seager}, D.~{Deming},
  K.~{Menou}, J.~{Cho}, L.J. {Richardson}, Science \textbf{27}, 623 (2006)

\bibitem{Crossfield2010}
I.J.M. {Crossfield}, B.M.S. {Hansen}, J.~{Harrington}, J.~{Cho}, D.~{Deming},
  K.~{Menou}, S.~{Seager}, ApJ \textbf{723}, 1436 (2010).
\newblock \doi{10.1088/0004-637X/723/2/1436}

\bibitem{Laughlin2009}
G.~{Laughlin}, D.~{Deming}, J.~{Langton}, D.~{Kasen}, S.~{Vogt}, P.~{Butler},
  E.~{Rivera}, S.~{Meschiari}, Nature \textbf{457}, 562 (2009)

\bibitem{Kipping2010}
D.M. {Kipping}, S.J. {Fossey}, G.~{Campanella}, MNRAS \textbf{400}, 398 (2009).
\newblock \doi{10.1111/j.1365-2966.2009.15472.x}

\bibitem{Segura2005}
A.~{Segura}, J.F. {Kasting}, V.~{Meadows}, M.~{Cohen}, J.~{Scalo}, D.~{Crisp},
  R.A.H. {Butler}, G.~{Tinetti}, Astrobiology \textbf{5}, 706 (2005).
\newblock \doi{10.1089/ast.2005.5.706}

\bibitem{Grenfell2011a}
J.L. {Grenfell}, S.~{Gebauer}, K.~{Palczynski}, R.~{Rauer}, H.and~{Lehmann},
  P.~{von Paris}, J.~{Cabrera}, M.~{Godolt}, F.~{Belu}, A.and~{Selsis},
  P.~{Hedelt}, A\&A p. submitted (2011)

\bibitem{TinettiIAU}
G.~{Tinetti}, J.Y.K. {Cho}, C.A. {Griffith}, O.~{Grasset}, L.~{Grenfell},
  T.~{Guillot}, T.T. {Koskinen}, J.I. {Moses}, D.~{Pinfield}, J.~{Tennyson},
  M.~{Tessenyi}, R.~{Wordsworth}, A.~{Aylward}, R.~{van Boekel}, A.~{Coradini},
  T.~{Encrenaz}, I.~{Snellen}, M.R. {Zapatero-Osorio}, J.~{Bouwman}, V.C. {du
  Foresto}, M.~{Lopez-Morales}, I.~{Mueller-Wodarg}, E.~{Pall{\'e}},
  F.~{Selsis}, A.~{Sozzetti}, J.P. {Beaulieu}, T.~{Henning}, M.~{Meyer},
  G.~{Micela}, I.~{Ribas}, D.~{Stam}, M.~{Swain}, O.~{Krause}, M.~{Ollivier},
  E.~{Pace}, B.~{Swinyard}, P.A.R. {Ade}, N.~{Achilleos}, A.~{Adriani}, C.B.
  {Agnor}, C.~{Afonso}, C.A. {Prieto}, G.~{Bakos}, R.J. {Barber}, M.~{Barlow},
  P.~{Bernath}, B.~{B{\'e}zard}, P.~{Bord{\'e}}, L.R. {Brown}, A.~{Cassan},
  C.~{Cavarroc}, A.~{Ciaravella}, C.~{Cockell}, A.~{Coustenis}, C.~{Danielski},
  L.~{Decin}, R.~{De Kok}, O.~{Demangeon}, P.~{Deroo}, P.~{Doel},
  P.~{Drossart}, L.N. {Fletcher}, M.~{Focardi}, F.~{Forget}, S.~{Fossey},
  P.~{Fouqu{\'e}}, J.~{Frith}, M.~{Galand}, P.~{Gaulme}, J.I.G.
  {Hern{\'a}ndez}, D.~{Grassi}, M.J. {Griffin}, U.~{Gr{\"o}zinger},
  M.~{Guedel}, P.~{Guio}, O.~{Hainaut}, R.~{Hargreaves}, P.H. {Hauschildt},
  K.~{Heng}, D.~{Heyrovsky}, R.~{Hueso}, P.~{Irwin}, L.~{Kaltenegger},
  P.~{Kervella}, D.~{Kipping}, G.~{Kovacs}, A.L. {Barbera}, H.~{Lammer},
  E.~{Lellouch}, G.~{Leto}, M.L. {Morales}, M.A.L. {Valverde},
  M.~{Lopez-Puertas}, C.~{Lovi}, A.~{Maggio}, J.P. {Maillard}, J.M. {Prado},
  J.B. {Marquette}, F.J. {Martin-Torres}, P.~{Maxted}, S.~{Miller},
  S.~{Molinari}, D.~{Montes}, A.~{Moro-Martin}, O.~{Mousis}, N.N. {Tuong},
  R.~{Nelson}, G.S. {Orton}, E.~{Pantin}, E.~{Pascale}, S.~{Pezzuto},
  E.~{Poretti}, R.~{Prinja}, L.~{Prisinzano}, J.M. {R{\'e}ess}, A.~{Reiners},
  B.~{Samuel}, J.S. {Forcada}, D.~{Sasselov}, G.~{Savini}, B.~{Sicardy},
  A.~{Smith}, L.~{Stixrude}, G.~{Strazzulla}, G.~{Vasisht}, S.~{Vinatier},
  S.~{Viti}, I.~{Waldmann}, G.J. {White}, T.~{Widemann}, R.~{Yelle}, Y.~{Yung},
  S.~{Yurchenko}, in \emph{IAU Symposium}, \emph{IAU Symposium}, vol. 276, ed.
  by {A.~Sozzetti, M.~G.~Lattanzi, \& A.~P.~Boss} (2011), \emph{IAU Symposium},
  vol. 276, pp. 359--370.
\newblock \doi{10.1017/S1743921311020448}

\bibitem{schneider}
J.~Schneider,   (2011).
\newblock \urlprefix\url{{http://www.exoplanet.eu}}

\bibitem{Leger2009}
A.~{Leger}, D.~{Rouan}, J.~{Schneider}, P.~{Barge}, M.~{Fridlund}, B.~{Samuel},
  M.~{Ollivier}, E.~{Guenther}, M.~{Deleuil}, H.~{Deeg}, M.~{Auvergne},
  R.~{Alonso}, S.~{Aigrain}, A.~{Alapini}, J.~{Almenara}, A.~{Baglin},
  M.~{Barbieri}, H.~{Bruntt}, P.~{Borde}, F.~{Bouchy}, J.~{Cabrera},
  C.~{Catala}, L.~{Carone}, S.~{Carpano}, S.~{Csizmadia}, R.~{Dvorak},
  A.~{Erikson}, S.~{Ferraz-Mello}, B.~{Foing}, F.~{Fressin}, D.~{Gandolfi},
  M.~{Gillon}, P.~{Gondoin}, O.~{Grasset}, T.~{Guillot}, A.~{Hatzes},
  G.~{Hebrard}, L.~{Jorda}, H.~{Lammer}, A.~{Llebaria}, B.~{Loeillet},
  M.~{Mayor}, T.~{Mazeh}, C.~{Moutou}, M.~{Paetzold}, F.~{Pont}, D.~{Queloz},
  H.~{Rauer}, S.~{Renner}, R.~{Samadi}, A.~{Shporer}, C.~{Sotin}, B.~{Tingley},
  G.~{Wuchterl}, A\&A \textbf{506}, 287 (2009)

\bibitem{Batalha2011}
N.M. {Batalha}, W.J. {Borucki}, S.T. {Bryson}, L.A. {Buchhave}, D.A.
  {Caldwell}, J.~{Christensen-Dalsgaard}, D.~{Ciardi}, E.W. {Dunham},
  F.~{Fressin}, T.N. {Gautier}, III, R.L. {Gilliland}, M.R. {Haas}, S.B.
  {Howell}, J.M. {Jenkins}, H.~{Kjeldsen}, D.G. {Koch}, D.W. {Latham}, J.J.
  {Lissauer}, G.W. {Marcy}, J.F. {Rowe}, D.D. {Sasselov}, S.~{Seager}, J.H.
  {Steffen}, G.~{Torres}, G.S. {Basri}, T.M. {Brown}, D.~{Charbonneau},
  J.~{Christiansen}, B.~{Clarke}, W.D. {Cochran}, A.~{Dupree}, D.C. {Fabrycky},
  D.~{Fischer}, E.B. {Ford}, J.~{Fortney}, F.R. {Girouard}, M.J. {Holman},
  J.~{Johnson}, H.~{Isaacson}, T.C. {Klaus}, P.~{Machalek}, A.V. {Moorehead},
  R.C. {Morehead}, D.~{Ragozzine}, P.~{Tenenbaum}, J.~{Twicken}, S.~{Quinn},
  J.~{VanCleve}, L.M. {Walkowicz}, W.F. {Welsh}, E.~{Devore}, A.~{Gould}, ApJ
  \textbf{729}, 27 (2011).
\newblock \doi{10.1088/0004-637X/729/1/27}

\bibitem{Winn2011}
J.N. {Winn}, J.M. {Matthews}, R.I. {Dawson}, D.~{Fabrycky}, M.J. {Holman},
  T.~{Kallinger}, R.~{Kuschnig}, D.~{Sasselov}, D.~{Dragomir}, D.B. {Guenther},
  A.F.J. {Moffat}, J.F. {Rowe}, S.~{Rucinski}, W.W. {Weiss}, ApJL \textbf{737},
  L18+ (2011).
\newblock \doi{10.1088/2041-8205/737/1/L18}

\bibitem{Charbonneau2010}
D.~{Charbonneau}, Z.K. {Berta}, J.~{Irwin}, C.J. {Burke}, P.~{Nutzman}, L.A.
  {Buchhave}, C.~{Lovis}, X.~{Bonfils}, D.W. {Latham}, S.~{Udry}, R.A.
  {Murray-Clay}, M.J. {Holman}, E.E. {Falco}, J.N. {Winn}, D.~{Queloz},
  F.~{Pepe}, M.~{Mayor}, X.~{Delfosse}, T.~{Forveille}, Nature \textbf{462},
  891 (2009)

\bibitem{Henry2011}
G.W. {Henry}, A.W. {Howard}, G.W. {Marcy}, D.A. {Fischer}, J.A. {Johnson},
  ArXiv e-prints  (2011)

\bibitem{Marcy1996}
G.W. {Marcy}, R.P. {Butler}, ApJ \textbf{464}, L147+ (1996)

\bibitem{Rivera2005}
E.J. {Rivera}, J.J. {Lissauer}, R.P. {Butler}, G.W. {Marcy}, S.S. {Vogt}, D.A.
  {Fischer}, T.M. {Brown}, G.~{Laughlin}, G.W. {Henry}, ApJ \textbf{634}, 625
  (2005).
\newblock \doi{10.1086/491669}

\bibitem{Mayor2009}
M.~{Mayor}, X.~{Bonfils}, T.~{Forveille}, X.~{Delfosse}, S.~{Udry}, J.L.
  {Bertaux}, H.~{Beust}, F.~{Bouchy}, C.~{Lovis}, F.~{Pepe}, C.~{Perrier},
  D.~{Queloz}, N.C. {Santos}, A\&A \textbf{507}, 487 (2009)

\bibitem{Mayor2011}
M.~{Mayor}, M.~{Marmier}, C.~{Lovis}, S.~{Udry}, D.~{S{\'e}gransan}, F.~{Pepe},
  W.~{Benz}, J.. {Bertaux}, F.~{Bouchy}, X.~{Dumusque}, G.~{Lo Curto},
  C.~{Mordasini}, D.~{Queloz}, N.C. {Santos}, ArXiv e-prints  (2011)

\bibitem{Borucki2011}
W.J. {Borucki}, D.G. {Koch}, G.~{Basri}, N.~{Batalha}, T.M. {Brown}, S.T.
  {Bryson}, D.~{Caldwell}, J.~{Christensen-Dalsgaard}, W.D. {Cochran},
  E.~{DeVore}, E.W. {Dunham}, T.N. {Gautier}, III, J.C. {Geary},
  R.~{Gilliland}, A.~{Gould}, S.B. {Howell}, J.M. {Jenkins}, D.W. {Latham},
  J.J. {Lissauer}, G.W. {Marcy}, J.~{Rowe}, D.~{Sasselov}, A.~{Boss},
  D.~{Charbonneau}, D.~{Ciardi}, L.~{Doyle}, A.K. {Dupree}, E.B. {Ford},
  J.~{Fortney}, M.J. {Holman}, S.~{Seager}, J.H. {Steffen}, J.~{Tarter}, W.F.
  {Welsh}, C.~{Allen}, L.A. {Buchhave}, J.L. {Christiansen}, B.D. {Clarke},
  S.~{Das}, J.M. {D{\'e}sert}, M.~{Endl}, D.~{Fabrycky}, F.~{Fressin},
  M.~{Haas}, E.~{Horch}, A.~{Howard}, H.~{Isaacson}, H.~{Kjeldsen},
  J.~{Kolodziejczak}, C.~{Kulesa}, J.~{Li}, P.W. {Lucas}, P.~{Machalek},
  D.~{McCarthy}, P.~{MacQueen}, S.~{Meibom}, T.~{Miquel}, A.~{Prsa}, S.N.
  {Quinn}, E.V. {Quintana}, D.~{Ragozzine}, W.~{Sherry}, A.~{Shporer},
  P.~{Tenenbaum}, G.~{Torres}, J.D. {Twicken}, J.~{Van Cleve}, L.~{Walkowicz},
  F.C. {Witteborn}, M.~{Still}, ApJ \textbf{736}, 19 (2011).
\newblock \doi{10.1088/0004-637X/736/1/19}

\bibitem{Howard2011}
A.W. {Howard}, G.W. {Marcy}, S.T. {Bryson}, J.M. {Jenkins}, J.F. {Rowe}, N.M.
  {Batalha}, W.J. {Borucki}, D.G. {Koch}, E.W. {Dunham}, T.N. {Gautier}, III,
  J.~{Van Cleve}, W.D. {Cochran}, D.W. {Latham}, J.J. {Lissauer}, G.~{Torres},
  T.M. {Brown}, R.L. {Gilliland}, L.A. {Buchhave}, D.A. {Caldwell},
  J.~{Christensen-Dalsgaard}, D.~{Ciardi}, F.~{Fressin}, M.R. {Haas}, S.B.
  {Howell}, H.~{Kjeldsen}, S.~{Seager}, L.~{Rogers}, D.D. {Sasselov}, J.H.
  {Steffen}, G.S. {Basri}, D.~{Charbonneau}, J.~{Christiansen}, B.~{Clarke},
  A.~{Dupree}, D.C. {Fabrycky}, D.A. {Fischer}, E.B. {Ford}, J.J. {Fortney},
  J.~{Tarter}, F.R. {Girouard}, M.J. {Holman}, J.A. {Johnson}, T.C. {Klaus},
  P.~{Machalek}, A.V. {Moorhead}, R.C. {Morehead}, D.~{Ragozzine},
  P.~{Tenenbaum}, J.D. {Twicken}, S.N. {Quinn}, H.~{Isaacson}, A.~{Shporer},
  P.W. {Lucas}, L.M. {Walkowicz}, W.F. {Welsh}, A.~{Boss}, E.~{Devore},
  A.~{Gould}, J.C. {Smith}, R.L. {Morris}, A.~{Prsa}, T.D. {Morton}, ArXiv
  e-prints  (2011)

\bibitem{Cassan2011}
A.~{Cassan}, D.~{Kubas}, J.P. {Beaulieu}, et~al., Nature p. in press (2011)

\bibitem{Tessenyi2011}
M.~Tessenyi, M.~Ollivier, G.~Tinetti, J.P. Beaulieu, V.~{Coude du Foresto},
  T.~Encrenaz, G.~Micela, B.~Swinyard, I.~Ribas, A.~Aylward, J.~Tennyson, M.R.
  Swain, A, Sozzetti, G.~Vasisht, P.~Deroo, ApJ p. in press (2011)

\bibitem{Lepine2011}
S.~{L{\'e}pine}, E.~{Gaidos}, AJ \textbf{142}, 138 (2011).
\newblock \doi{10.1088/0004-6256/142/4/138}

\bibitem{Mandell2007}
A.M. {Mandell}, S.N. {Raymond}, S.~{Sigurdsson}, ApJ \textbf{660}, 823 (2007).
\newblock \doi{10.1086/512759}

\bibitem{Borucki2009}
W.J. {Borucki}, D.~{Koch}, J.~{Jenkins}, D.~{Sasselov}, R.~{Gilliland},
  N.~{Batalha}, D.W. {Latham}, D.~{Caldwell}, G.~{Basri}, T.~{Brown},
  J.~{Christensen-Dalsgaard}, W.D. {Cochran}, E.~{DeVore}, E.~{Dunham}, A.K.
  {Dupree}, T.~{Gautier}, J.~{Geary}, A.~{Gould}, S.~{Howell}, H.~{Kjeldsen},
  J.~{Lissauer}, G.~{Marcy}, S.~{Meibom}, D.~{Morrison}, J.~{Tarter}, Science
  \textbf{325}, 709 (2009).
\newblock \doi{10.1126/science.1178312}

\bibitem{Seager2000}
S.~{Seager}, D.D. {Sasselov}, ApJ \textbf{537}, 916 (2000).
\newblock \doi{10.1086/309088}

\bibitem{brown}
T.M. {Brown}, ApJ \textbf{553}, 1006 (2001)

\bibitem{tinettia}
G.~{Tinetti}, M.C. {Liang}, A.~{Vidal-Madjar}, D.~{Ehrenreich}, A.L. des
  Etangs, Y.~{Yung}, ApJ \textbf{654}, L99 (2007)

\bibitem{charbonneau2}
D.~{Charbonneau}, T.M. {Brown}, R.W. {Noyes}, R.L. {Gilliland}, ApJ
  \textbf{568}, 377 (2002)

\bibitem{VidalMadjar2003}
A.~{Vidal-Madjar}, A.~{Lecavelier des Etangs}, J.M. {D{\'e}sert}, G.E.
  {Ballester}, R.~{Ferlet}, G.~{H{\'e}brard}, M.~{Mayor}, Nature \textbf{422},
  143 (2003).
\newblock \doi{10.1038/nature01448}

\bibitem{Knutson2007}
H.A. {Knutson}, D.~{Charbonneau}, R.W. {Noyes}, T.M. {Brown}, R.L. {Gilliland},
  ApJ \textbf{655}, 564 (2007).
\newblock \doi{10.1086/510111}

\bibitem{Pont2008}
F.~{Pont}, H.~{Knutson}, R.L. {Gilliland}, C.~{Moutou}, D.~{Charbonneau}, MNRAS
  \textbf{385}, 109 (2008).
\newblock \doi{10.1111/j.1365-2966.2008.12852.x}

\bibitem{redfeild}
S.~{Redfield}, M.~{Endl}, W.~{Cochran}, L.~{Koesterke}, ApJ \textbf{673}, 87
  (2008)

\bibitem{snellen}
I.~{Snellen}, S.~{Albrecht}, E.~{de Mooij}, R.~{Le Poole}, A\&A \textbf{487},
  357 (2008)

\bibitem{bean2010}
J.L. {Bean}, E.M.R. {Kempton}, D.~{Homeier}, Nature \textbf{468}, 669 (2010)

\bibitem{Knutson2007a}
H.A. {Knutson}, D.~{Charbonneau}, L.E. {Allen}, J.J. {Fortney}, E.~{Agol}, N.B.
  {Cowan}, A.P. {Showman}, C.S. {Cooper}, S.T. {Megeath}, Nature \textbf{447},
  183 (2007).
\newblock \doi{10.1038/nature05782}

\bibitem{tinettib}
G.~{Tinetti}, A.~{Vidal-Madjar}, M.C. {Liang}, J.P. {Beaulieu}, Y.~{Yung},
  S.~{Carey}, R.J. {Barber}, J.~{Tennyson}, I.~{Ribas}, N.~{Allard}, G.E.
  {Ballester}, D.K. {Sing}, F.~{Selsis}, Nature \textbf{448}, 169 (2007)

\bibitem{tinettic}
G.~{Tinetti}, P.~{Deroo}, M.R. {Swain}, C.A. {Griffith}, G.~{Vasisht}, L.R.
  {Brown}, C.~{Burke}, P.~{McCullough}, ApJ \textbf{712}, L139 (2010)

\bibitem{swain2008a}
M.R. {Swain}, G.~{Vasisht}, G.~{Tinetti}, Nature \textbf{452}, 329 (2008)

\bibitem{beaulieu}
J.P. {Beaulieu}, S.~{Carey}, I.~{Ribas}, G.~{Tinetti}, ApJ \textbf{677}, 1343
  (2008)

\bibitem{beaulieu2010a}
J.P. {Beaulieu}, D.M. {Kipping}, V.~{Batista}, G.~{Tinetti}, I.~{Ribas},
  S.~{Carey}, J.A. Noriega-Crespo, C.A. {Griffith}, G.~{Campanella}, S.~{Dong},
  J.~{Tennyson}, R.J. {Barber}, P.~{Deroo}, S.J. {Fossey}, D.~{Liang}, M.R.
  {Swain}, Y.~{Yung}, N.~{Allard}, MNRAS \textbf{409}, 369 (2010)

\bibitem{Beaulieu2010b}
J.P. {Beaulieu}, G.~{Tinetti}, D.M. {Kipping}, I.~{Ribas}, R.J. {Barber}, J.K.
  {Cho}, I.~{Polichtchouk}, J.~{Tennyson}, S.N. {Yurchenko}, C.A. {Griffith},
  V.~{Batista}, I.~{Waldmann}, S.~{Miller}, S.~{Carey}, O.~{Mousis}, S.J.
  {Fossey}, A.~{Aylward}, ArXiv e-prints  (2010)

\bibitem{Agol2010}
E.~{Agol}, N.B. {Cowan}, H.A. {Knutson}, D.~{Deming}, J.H. {Steffen}, G.W.
  {Henry}, D.~{Charbonneau}, ApJ \textbf{721}, 1861 (2010).
\newblock \doi{10.1088/0004-637X/721/2/1861}

\bibitem{deming}
D.~{Deming}, S.~{Seager}, L.J. {Richardson}, J.~{Harrington}, Nature
  \textbf{434}, 740 (2005)

\bibitem{charbonneau3}
D.~{Charbonneau}, L.E. {Allen}, G.~S.~Thomas Megeath~and{Torres}, R.~{Alonso},
  T.M. {Brown}, R.L. {Gilliland}, D.W. {Latham}, G.~{Mandushev}, F.T.
  {O'Donovan}, A.~{Sozzetti}, ApJ \textbf{626}, 523 (2005)

\bibitem{swain2009a}
M.R. {Swain}, G.~{Vasisht}, G.~{Tinetti}, J.~{Bouwman}, Y.~Pin Chen~and{Yung},
  D.~{Deming}, P.~{Deroo}, ApJ \textbf{960}, L114 (2009)

\bibitem{swain2009b}
M.R. {Swain}, G.~{Tinetti}, G.~{Vasisht}, P.~{Deroo}, C.~{Griffith},
  J.~{Bouwman}, Y.~Pin Chen~and{Yung}, A.~{Burrows}, L.~{Brown}, J.~{Matthews},
  J.F. {Roe}, R.~{Kuschnig}, D.~{Angerhausen}, ApJ \textbf{704}, 1616 (2009)

\bibitem{swain2010}
M.R. {Swain}, P.~{Deroo}, C.A. {Griffith}, G.~{Tinetti}, A.~{Thatte},
  G.~{Vasishtand}, P.~{Chen}, J.~{Bouwman}, I.J. {Crossfield},
  D.~{Angerhausen}, C.~{Afonso}, T.~{Henning}, Nature \textbf{463}, 637 (2010)

\bibitem{Stevenson2010}
K.B. {Stevenson}, J.~{Harrington}, S.~{Nymeyer}, N.~{Madhusudhan}, S.~{Seager},
  W.C. {Bowman}, R.A. {Hardy}, D.~{Deming}, E.~{Rauscher}, N.B. {Lust}, Nature
  \textbf{464}, 1161 (2010).
\newblock \doi{10.1038/nature09013}

\bibitem{Rowe2006}
J.F. {Rowe}, J.M. {Matthews}, S.~{Seager}, R.~{Kuschnig}, D.B. {Guenther},
  A.F.J. {Moffat}, S.M. {Rucinski}, D.~{Sasselov}, G.A.H. {Walker}, W.W.
  {Weiss}, ApJ \textbf{646}, 1241 (2006).
\newblock \doi{10.1086/504252}

\bibitem{Kipping2011}
D.M. {Kipping}, D.S. {Spiegel}, MNRAS  (2011)

\bibitem{Snellen2009}
I.A.G. {Snellen}, E.J.W. {de Mooij}, S.~{Albrecht}, Nature \textbf{459}, 543
  (2009).
\newblock \doi{10.1038/nature08045}

\bibitem{KippingTinetti2010}
D.M. {Kipping}, G.~{Tinetti}, MNRAS \textbf{407}, 2589 (2010).
\newblock \doi{10.1111/j.1365-2966.2010.17094.x}

\bibitem{Rauscher2007}
E.~{Rauscher}, K.~{Menou}, J.~{Cho}, S.~{Seager}, B.M.S. {Hansen}, ApJ
  \textbf{662}, L115 (2007).
\newblock \doi{10.1086/519374}

\bibitem{Cowan2009}
N.B. {Cowan}, E.~{Agol}, V.S. {Meadows}, T.~{Robinson}, T.A. {Livengood},
  D.~{Deming}, C.M. {Lisse}, M.F. {A'Hearn}, D.D. {Wellnitz}, S.~{Seager},
  D.~{Charbonneau}, {the EPOXI Team}, ApJ \textbf{700}, 915 (2009).
\newblock \doi{10.1088/0004-637X/700/2/915}

\bibitem{Goody1989}
R.M. {Goody}, Y.L. {Yung}, \emph{{Atmospheric radiation : theoretical basis}}
  (1989)

\bibitem{Hanel}
R.A. {Hanel}, B.J. {Conrath}, D.~{Jennings}, R.E. {Samuelson}, B.J. {Conrath},
  D.~{Jennings}, R.E. {Samuelson}, \emph{{Book Review: Exploration of the solar
  system by infrared remote sensing / Cambridge U Press, 1992}}, vol. 102
  (1992)

\bibitem{madhu}
N.~{Madhusudhan}, S.~{Seager}, ApJ \textbf{707}, 24 (2009).
\newblock \doi{10.1088/0004-637X/707/1/24}

\bibitem{TinettiFaraday}
G.~{Tinetti}, C.A. {Griffith}, M.R. {Swain}, P.~{Deroo}, J.P. {Beaulieu},
  G.~{Vasisht}, D.~{Kipping}, I.~{Waldmann}, J.~{Tennyson}, R.J. {Barber},
  J.~{Bouwman}, N.~{Allard}, L.R. {Brown}, Faraday Discussions \textbf{147},
  369 (2010).
\newblock \doi{10.1039/c005126h}

\bibitem{Lee2011}
J.M. {Lee}, L.N. {Fletcher}, P.G.J. {Irwin}, MNRAS p. 1983 (2011).
\newblock \doi{10.1111/j.1365-2966.2011.20013.x}

\bibitem{Line2011}
M.R. {Line}, X.~{Zhang}, G.~{Vaisht}, P.~{Chen}, V.~{Natraj}, P.~{Chen}, Y.L.
  {Yung}, ApJL  (2011)

\bibitem{Walkowicz2009}
L.M. {Walkowicz}, S.L. {Hawley}, in \emph{American Institute of Physics
  Conference Series}, vol. 1094, ed. by {E.~Stempels} (2009), vol. 1094, pp.
  696--699.
\newblock \doi{10.1063/1.3099209}

\bibitem{Basri2011}
G.~{Basri}, L.M. {Walkowicz}, N.~{Batalha}, R.L. {Gilliland}, J.~{Jenkins},
  W.J. {Borucki}, D.~{Koch}, D.~{Caldwell}, A.K. {Dupree}, D.W. {Latham}, G.W.
  {Marcy}, S.~{Meibom}, T.~{Brown}, AJ \textbf{141}, 20 (2011).
\newblock \doi{10.1088/0004-6256/141/1/20}

\bibitem{Casertano2008}
S.~{Casertano}, M.G. {Lattanzi}, A.~{Sozzetti}, A.~{Spagna}, S.~{Jancart},
  R.~{Morbidelli}, R.~{Pannunzio}, D.~{Pourbaix}, D.~{Queloz}, A\&A
  \textbf{482}, 699 (2008).
\newblock \doi{10.1051/0004-6361:20078997}

\bibitem{Sozzetti2011}
A.~{Sozzetti}, in \emph{EAS Publications Series}, \emph{EAS Publications
  Series}, vol.~45 (2011), \emph{EAS Publications Series}, vol.~45, pp.
  273--278.
\newblock \doi{10.1051/eas/1045046}

\bibitem{Bakos2002}
G.{\'A}. {Bakos}, J.~{L{\'a}z{\'a}r}, I.~{Papp}, P.~{S{\'a}ri}, E.M. {Green},
  PASP \textbf{114}, 974 (2002).
\newblock \doi{10.1086/342382}

\bibitem{Bakos2009}
G.~{Bakos}, C.~{Afonso}, T.~{Henning}, A.~{Jord{\'a}n}, M.~{Holman}, R.W.
  {Noyes}, P.D. {Sackett}, D.~{Sasselov}, G.~{Kov{\'a}cs}, Z.~{Csubry},
  A.~{P{\'a}l}, in \emph{IAU Symposium}, \emph{IAU Symposium}, vol. 253 (2009),
  \emph{IAU Symposium}, vol. 253, pp. 354--357.
\newblock \doi{10.1017/S174392130802663X}

\bibitem{Pollacco2006}
D.L. {Pollacco}, I.~{Skillen}, A.~{Collier Cameron}, D.J. {Christian},
  C.~{Hellier}, J.~{Irwin}, T.A. {Lister}, R.A. {Street}, R.G. {West},
  D.~{Anderson}, W.I. {Clarkson}, H.~{Deeg}, B.~{Enoch}, A.~{Evans},
  A.~{Fitzsimmons}, C.A. {Haswell}, S.~{Hodgkin}, K.~{Horne}, S.R. {Kane}, F.P.
  {Keenan}, P.F.L. {Maxted}, A.J. {Norton}, J.~{Osborne}, N.R. {Parley}, R.S.I.
  {Ryans}, B.~{Smalley}, P.J. {Wheatley}, D.M. {Wilson}, PASP \textbf{118},
  1407 (2006).
\newblock \doi{10.1086/508556}

\bibitem{XO}
P.R. {McCullough}, J.E. {Stys}, J.A. {Valenti}, S.W. {Fleming}, K.A. {Janes},
  J.N. {Heasley}, PASP \textbf{117}, 783 (2005).
\newblock \doi{10.1086/432024}

\bibitem{Gillon2007}
M.~{Gillon}, F.~{Pont}, B.~{Demory}, F.~{Mallmann}, M.~{Mayor}, T.~{Mazeh},
  D.~{Queloz}, A.~{Shporer}, S.~{Udry}, C.~{Vuissoz}, A\&A \textbf{472}, L13
  (2007).
\newblock \doi{10.1051/0004-6361:20077799}

\bibitem{Nutzman2008}
P.~{Nutzman}, D.~{Charbonneau}, PASP \textbf{120}, 317 (2008).
\newblock \doi{10.1086/533420}

\bibitem{Apache}
A.~{Sozzetti}, M.G. {Lattanzi}, S.~{Ligori}, R.L. {Smart}, {others}, in
  \emph{Pathways Towards Habitable Planets}, \emph{Internal report}, vol. 110,
  ed. by {Publications OATO} (2008), \emph{Internal report}, vol. 110, pp.
  20--+

\bibitem{Lovis2009}
C.~{Lovis}, M.~{Mayor}, F.~{Bouchy}, F.~{Pepe}, D.~{Queloz}, S.~{Udry},
  W.~{Benz}, C.~{Mordasini}, in \emph{IAU Symposium}, vol. 253 (2009), vol.
  253, pp. 502--505.
\newblock \doi{10.1017/S1743921308027051}

\bibitem{Quirrenbach}
A.~{Quirrenbach}, P.J. {Amado}, H.~{Mandel}, J.A. {Caballero}, I.~{Ribas},
  A.~{Reiners}, R.~{Mundt}, {CARMENES Consortium}, in \emph{Pathways Towards
  Habitable Planets}, \emph{Astronomical Society of the Pacific Conference
  Series}, vol. 430, ed. by {V.~Coud{\'e} Du Foresto, D.~M.~Gelino, \&
  I.~Ribas} (2010), \emph{Astronomical Society of the Pacific Conference
  Series}, vol. 430, pp. 521--+

\bibitem{Pepe2010}
F.A. {Pepe}, S.~{Cristiani}, R.~{Rebolo Lopez}, N.C. {Santos}, A.~{Amorim},
  G.~{Avila}, W.~{Benz}, P.~{Bonifacio}, A.~{Cabral}, P.~{Carvas}, R.~{Cirami},
  J.~{Coelho}, M.~{Comari}, I.~{Coretti}, V.~{de Caprio}, H.~{Dekker},
  B.~{Delabre}, P.~{di Marcantonio}, V.~{D'Odorico}, M.~{Fleury},
  R.~{Garc{\'{\i}}a}, J.M. {Herreros Linares}, I.~{Hughes}, O.~{Iwert},
  J.~{Lima}, J.L. {Lizon}, G.~{Lo Curto}, C.~{Lovis}, A.~{Manescau},
  C.~{Martins}, D.~{M{\'e}gevand}, A.~{Moitinho}, P.~{Molaro}, M.~{Monteiro},
  M.~{Monteiro}, L.~{Pasquini}, C.~{Mordasini}, D.~{Queloz}, J.L. {Rasilla},
  J.M. {Rebord{\~a}o}, S.~{Santana Tschudi}, P.~{Santin}, D.~{Sosnowska},
  P.~{Span{\`o}}, F.~{Tenegi}, S.~{Udry}, E.~{Vanzella}, M.~{Viel}, M.R.
  {Zapatero Osorio}, F.~{Zerbi}, in \emph{Society of Photo-Optical
  Instrumentation Engineers (SPIE) Conference Series}, \emph{Society of
  Photo-Optical Instrumentation Engineers (SPIE) Conference Series}, vol. 7735
  (2010), \emph{Society of Photo-Optical Instrumentation Engineers (SPIE)
  Conference Series}, vol. 7735.
\newblock \doi{10.1117/12.857122}

\bibitem{Corot2011}
M.~{Deleuil}, C.~{Moutou}, P.~{Bord{\'e}}, Detection and Dynamics of Transiting
  Exoplanets, St.~Michel l'Observatoire, France, Edited by F.~Bouchy;
  R.~D{\'{\i}}az; C.~Moutou; EPJ Web of Conferences, Volume 11, id.01001
  \textbf{11}, 01001 (2011).
\newblock \doi{10.1051/epjconf/20101101001}

\bibitem{TESS}
G.R. {Ricker}, D.W. {Latham}, R.K. {Vanderspek}, K.A. {Ennico}, G.~{Bakos},
  T.M. {Brown}, A.J. {Burgasser}, D.~{Charbonneau}, L.D. {Deming}, J.P. {Doty},
  E.W. {Dunham}, J.L. {Elliot}, M.J. {Holman}, S.~{Ida}, J.M. {Jenkins}, J.G.
  {Jernigan}, N.~{Kawai}, G.P. {Laughlin}, J.J. {Lissauer}, F.~{Martel}, D.D.
  {Sasselov}, R.H. {Schingler}, S.~{Seager}, G.~{Torres}, S.~{Udry}, J.S.
  {Villasenor}, J.N. {Winn}, S.P. {Worden}, in \emph{American Astronomical
  Society Meeting Abstracts \#213}, \emph{Bulletin of the American Astronomical
  Society}, vol.~41 (2009), \emph{Bulletin of the American Astronomical
  Society}, vol.~41, p. 403.01

\bibitem{PLATO}
C.~{Catala}, {PLATO Consortium}, Journal of Physics Conference Series
  \textbf{118}(1), 012040 (2008).
\newblock \doi{10.1088/1742-6596/118/1/012040}

\bibitem{Pilbratt2010}
G.L. {Pilbratt}, J.R. {Riedinger}, T.~{Passvogel}, G.~{Crone}, D.~{Doyle},
  U.~{Gageur}, A.M. {Heras}, C.~{Jewell}, L.~{Metcalfe}, S.~{Ott},
  M.~{Schmidt}, A\&A \textbf{518}, L1+ (2010).
\newblock \doi{10.1051/0004-6361/201014759}

\bibitem{Terrier2010}
B.~{Terrier}, A.~{Delannoy}, P.~{Chorier}, M.~{Maillard}, L.~{Rubaldo}, in
  \emph{Society of Photo-Optical Instrumentation Engineers (SPIE) Conference
  Series}, \emph{Society of Photo-Optical Instrumentation Engineers (SPIE)
  Conference Series}, vol. 7826 (2010), \emph{Society of Photo-Optical
  Instrumentation Engineers (SPIE) Conference Series}, vol. 7826.
\newblock \doi{10.1117/12.865053}

\bibitem{Itsuno2011}
A.M. {Itsuno}, J.D. {Phillips}, S.~{Velicu}, Journal of Electronic Materials
  \textbf{40}, 1624 (2011).
\newblock \doi{10.1007/s11664-011-1614-0}

\bibitem{VanCleve1995}
J.E. {Van Cleve}, T.L. {Herter}, R.~{Butturini}, G.E. {Gull}, J.R. {Houck},
  B.~{Pirger}, J.~{Schoenwald}, in \emph{Society of Photo-Optical
  Instrumentation Engineers (SPIE) Conference Series}, \emph{Society of
  Photo-Optical Instrumentation Engineers (SPIE) Conference Series}, vol. 2553,
  ed. by {M.~S.~Scholl \& B.~F.~Andresen} (1995), \emph{Society of
  Photo-Optical Instrumentation Engineers (SPIE) Conference Series}, vol. 2553,
  pp. 502--513

\bibitem{Planck}
{Planck Collaboration}, P.A.R. {Ade}, N.~{Aghanim}, M.~{Arnaud}, M.~{Ashdown},
  J.~{Aumont}, C.~{Baccigalupi}, M.~{Baker}, A.~{Balbi}, A.J. {Banday}, et~al.,
  ArXiv e-prints  (2011)

\end{thebibliography}
